\documentclass[article,nojss]{jss}

\usepackage{thumbpdf}
\usepackage{amsfonts,amstext,amsmath,amssymb,amsthm}
\usepackage{accents}
\usepackage{color}
\usepackage{rotating}
\usepackage{verbatim}
\usepackage{xspace}

\usepackage{booktabs}
\usepackage{dcolumn}
\usepackage[commandnameprefix=ifneeded,final]{changes}
\usepackage{hyphenat}

\newcommand{\SM}{Appendix}

\newcommand{\applications}{Section~\ref{sec:appl}}

\newcommand{\NPN}{NPN log-likelihood\xspace}
\newcommand{\sNPN}{smooth NPN log-likelihood\xspace}
\newcommand{\fNPN}{flow NPN log-likelihood\xspace}
\newcommand{\mNPN}{mixed NPN log-likelihood\xspace}

\makeatletter
\def\maxwidth{ %
  \ifdim\Gin@nat@width>\linewidth
    \linewidth
  \else
    \Gin@nat@width
  \fi
}
\makeatother

\definecolor{fgcolor}{rgb}{0.345, 0.345, 0.345}
\usepackage{framed}
\makeatletter
 {\par\unskip\endMakeFramed%
 \at@end@of@kframe}
\makeatother

\definecolor{shadecolor}{rgb}{.97, .97, .97}
\definecolor{messagecolor}{rgb}{0, 0, 0}
\definecolor{warningcolor}{rgb}{1, 0, 1}
\definecolor{errorcolor}{rgb}{1, 0, 0}
\newenvironment{knitrout}{}{} 

\usepackage{alltt}

\theoremstyle{example}

\newcommand{\pbox}{\mathcal{B}}

\usepackage{booktabs}
\usepackage{nicefrac}

\newcommand{\rk}{r}

\newcommand{\rY}{Y}
\newcommand{\rX}{\mX}
\newcommand{\rz}{z}
\newcommand{\ry}{y}
\newcommand{\rx}{\xvec}

\newcommand{\samY}{\mathcal{Y}}
\newcommand{\samX}{\mathcal{X}}

\newcommand{\h}{h}

\newcommand{\basisy}{\avec}

\newcommand{\parm}{\thetavec}
\newcommand{\eparm}{\theta}
\newcommand{\varparm}{\varthetavec}
\newcommand{\evarparm}{\vartheta}
\newcommand{\dimparm}{P}

\newcommand{\shiftparm}{\betavec}
\newcommand{\eshiftparm}{\beta}

\def \mA {\mLz_\text{A}}
\def \mB {\mLz_\text{B}}
\def \mC {\mLz_\text{C}}

\newcommand{\mLz}{\mOmega}

\newcommand{\ie}{\textit{i.e.}~}
\newcommand{\eg}{\textit{e.g.}~}

\renewcommand{\Prob}{\mathbb{P}}
\newcommand{\Ex}{\mathbb{E}}

\newcommand{\RR}{\mathbb{R}}
\newcommand{\NN}{\mathbb{N}}

\usepackage{dsfont}

 \DeclareMathOperator{\diag}{diag}

 \DeclareMathOperator{\ND}{N}
 
 \DeclareMathOperator{\UD}{U}

\def \avec {\text{\boldmath$a$}}   
\def \bvec {\text{\boldmath$b$}}   
\def \dvec {\text{\boldmath$d$}}    \def \mD {\text{\boldmath$D$}}
\def \evec {\text{\boldmath$e$}}

\def \hvec {\text{\boldmath$h$}}    
    \def \mI {\text{\boldmath$I$}}

    \def \mW {\text{\boldmath$W$}}
\def \xvec {\text{\boldmath$x$}}    \def \mX {\text{\boldmath$X$}}
\def \yvec {\text{\boldmath$y$}}    \def \mY {\text{\boldmath$Y$}}
\def \zvec {\text{\boldmath$z$}}    \def \mZ {\text{\boldmath$Z$}}

\def \betavec         {\text{\boldmath$\beta$}}
\def \gammavec        {\text{\boldmath$\gamma$}}

\def \etavec          {\text{\boldmath$\eta$}}
\def \thetavec        {\text{\boldmath$\theta$}}
\def \varthetavec     {\text{\boldmath$\vartheta$}}

\def \lambdavec       {\text{\boldmath$\lambda$}}
\def \muvec           {\text{\boldmath$\mu$}}

\def \upsilonvec      {\text{\boldmath$\upsilon$}}
\def \phivec          {\text{\boldmath$\phi$}}

\def \mLambda  {\mathbf{\Lambda}}

\def \mSigma   {\mathbf{\Sigma}}

\def \mPhi     {\mathbf{\Phi}}

\def \mOmega   {\mathbf{\Omega}}

\def \nullvec {\mathbf{0}}

\newtheorem{thm}{Theorem}
\newtheorem{coro}{Corollary}

\newcommand{\ubar}[1]{\underaccent{\bar}{#1}}

\author{Torsten Hothorn \\ Universit\"at Z\"urich}
\Plainauthor{Hothorn}

\title{On Nonparanormal Likelihoods}
\Plaintitle{On Nonparanormal Likelihoods}
\Shorttitle{On Nonparanormal Likelihoods}

\Abstract{

Nonparanormal models describe the joint distribution of multivariate responses via latent Gaussian, and thus parametric, copulae while allowing flexible nonparametric marginals. Some aspects of such distributions, for example conditional independence, are formulated parametrically. Other features, such as marginal distributions, can be formulated non- or semiparametrically. Such models are attractive when multivariate normality is questionable \added{but interpretability paramount}. 

Most estimation procedures perform two steps, first estimating the nonparametric part. The copula parameters come second, treating the marginal estimates as known. This is sufficient for some applications. For other applications, e.g.~when a semiparametric margin features parameters of interest or when standard errors are important, a simultaneous estimation of all parameters might be more advantageous.

\replaced{We present suitable parameterisations of nonparanormal models, possibly including semiparametric effects, and define four novel nonparanormal log-likelihood functions. In general, the corresponding one-step optimisation problems are shown to be non-convex. In some cases, however, biconvex problems emerge. Several convex approximations are discussed. From a low-level computational point of view, the core contribution is the score function for multivariate normal log-probabilities computed via Genz’ procedure.
}{We present suitable parameterisations of nonparanormal models, possibly including semiparametric effects, and define four novel nonparanormal log-likelihood functions. In general, the corresponding one-step optimisation problems are shown to be non-convex. In some cases, however, biconvex problems emerge. Several convex approximations are
discussed.}

\replaced{As a demonstration for the versatility of the theoretical and computational framework, we present a series of nonparanormal models for transformation discriminant analysis when some biomarkers are subject to limit-of-detection problems. Possible empirical gains of full maximum likelihood estimation compared to two-step approaches are illustrated in a simulation study targeting  semiparametric efficient polychoric correlation analysis where a theoretical benchmark is
available.}{From a low-level computational point of view, the core contribution is the score function for multivariate normal log-probabilities computed via Genz’ procedure. We present transformation discriminant analysis when some biomarkers are subject to limit-of-detection problems as an application and illustrate possible empirical gains in semiparametric efficient polychoric correlation
analysis.}
  }

\Plainkeywords{empirical likelihood, 
          censoring,
          Gaussian copula,
          mixed continuous-discrete responses,  
          normalising flows,
          transformation model}
\Address{
  Torsten Hothorn\\
  Institut f\"ur Epidemiologie, Biostatistik und Pr\"avention \\
  Universit\"at Z\"urich \\
  Hirschengraben 84, CH-8001 Z\"urich, Switzerland \\
  \texttt{Torsten.Hothorn@R-project.org}
}

\begin{document}


The multivariate normal distribution comes with a high potential for addiction
due to its covariance and precision matrix containing information about
marginal and conditional independence, respectively.  The fact that many
foundations of classical and contemporary multivariate statistics, such as
linear or quadratic discriminant analysis, graphical models, or structural
equation models, have been defined in terms of this distribution can be
explained by these favourable properties.
However, normality is the exception rather than the rule in the real world.  
As an alternative to a full
normality detox, statisticians may sacrifice marginal normality while retaining joint
normality on some latent scale.  This idea has been popularised under
different terms, for example as ``nonparanormal models''
\citep{Liu_Lafferty_Wasserman_2009} 
or ``coordinatewise Gaussianisation'' \citep{Mai_He_Zou_2023}, but its roots go
deeper.  For multivariate ordinal variables, \cite{Joergeskog_1994}
suggested the estimation of ``polychoric correlations'' defined by a latent
bivariate normal distribution coupled with marginal ordinal probit models. 
\added{Corresponding concepts in multivariate analysis are discussed in
\cite{Joe_1997} and} similar principles have been applied in semiparametric copula estimation
\citep{Klaassen_Wellner_1997, Joe_2005}, where marginal parameters are
estimated first, followed by a second step of estimating the copula
parameters \emph{conditionally} on margins.
The rank likelihood \citep{Hoff_2007,Hermes_Heerwaarden_Behrouzi_2024} 
does not
condition on marginal ranks but treats the marginal distributions as
nuisance parameters and focuses on the sole estimation of Gaussian copula parameters.

While such ideas have been very successfully applied for the estimation of
dependency structures also in high-dimensional multivariate
data, more complex models for both marginal and joint distributions are
necessary in many applications. Very much in the spirit of
\cite{Chen_Fan_Tsyrennikov_2006}, simultaneous likelihood estimation of and inference
for marginal \emph{and} copula parameters in more complex models is our main interest here.
The motivation comes from a wide range of applications of the
nonparanormal model where the application of a ``normalise and forget'' scheme is not
adequate. For general discrete \citep{Popovic_Hui_Warton_2018} or mixes of
continuous and discrete variables \citep[potentially allowing missing observations
in some responses,][]{Pritikon_Brick_Neale_2018,Christoffersen_2021,Goebler_2024},
rank-based approaches are more difficult to justify. Access to the full likelihood
covering all model parameters is required in semiparametric discriminant analysis
\citep{Mai_Zou_2014} or for analysing multivariate interval-censored survival data
\citep{Ding_Sun_2022}. Most interesting are nonparanormal models where
the marginal distributions, for example in multivariate regression models
\citep[\eg in multivariate GLMs or other linear models,][]{Lesaffre_Kaufmann_1992,Song_Li_Yuan_2009,Nikoloulopoulos_2023}, or the copula
parameters \citep[\eg in time-varying graphical models,][]{Lu_Kolar_Liu_2018}
feature parameters capturing covariate effects. The most striking example necessitating the
joint estimation of marginal and copula parameters is a model class for
survival analysis under dependent censoring
\citep{
Deresa_Keilegom_2024}. Here, marginally estimated
distributions for time-to-event and time-to-censoring are biased and only
the joint model leads to properly identified and estimable parameters. In
\applications, a selection of special applications and models with potential
benefits from the contributions to likelihood-based inference made in this work is discussed.

We proceed by suggesting parameterisations of the nonparanormal
model for discrete, continuous, and mixed discrete-continuous multivariate
responses and derive the \added{several variants of} nonparanormal
log-likelihoods, \added{including ``empirical'' or ``nonparametric'' and 
``parametric'' flavours,} and
the corresponding score functions. In general, maximum likelihood estimation in this model
class is shown to be non-convex. We discuss convex
approximations, which might be useful at least for the computation of
starting values. The theory and computational framework presented here
allows implementation of a rather general likelihood estimation toolbox for many interesting
applications. \added{For some areas in multivariate analysis discussed in
Section~\ref{sec:appl} and beyond, this toolbox helps to setup
likelihood and score functions. Coupled with suitable optimisation
routines, joint maximum likelihood estimation of all model parameters 
becomes computationally feasible in a range of interesting problems.
As an illustration highlighting the practical potential of this framework, 
we introduce novel forms of}
discriminant analyses for evaluating the diagnosis of hepatocellular carcinoma
based on \added{only} partially observed non-normal biomarker data.
\deleted{highlights the practical potential of this framework.}
\added{Regarding the statistical properties of maximum nonparanormal likelihood
estimation,} it is
demonstrated empirically that copula parameters obtained from optimising the nonparanormal
log-likelihood 
attain the semiparametric efficiency bound
derived by \cite{Klaassen_Wellner_1997}. \added{A more in-depth analysis of 
theoretical properties of these estimators is beyond the scope of this
paper.}

\section{The Nonparanormal Model} \label{sec:model}

We jointly observe $J$ response variables $\mY = (\rY_1, \dots, \rY_J)^\top$
from at least ordered sample spaces $\rY_j \in \samY_j, j = 1, \dots, J$.
The nonparanormal (NPN) model $\mY \sim \text{NPN}(\hvec, \mSigma)$ features
$J$ monotonically non-decreasing transformation functions 
$\hvec = (\h_1, \dots, \h_J)^\top$, one for each dimension
$\h_j: \samY_j \rightarrow \RR$ and, in addition, a positive
semidefinite $J \times J$ covariance matrix $\mSigma$ such that the joint
cumulative distribution function can be written in terms of normal
probabilities $\Prob(\mY \le \yvec) = \mPhi_\mSigma(\hvec(\yvec))$,
where $\mPhi_\mSigma$ is the joint cumulative distribution function of
$\ND_J(\nullvec, \mSigma)$. In case all elements of $\mY$ are 
continuous, each $\h_j$ is bijective and one typically
\citep{Liu_Lafferty_Wasserman_2009} writes $\text{NPN}(\hvec^{-1}, \mSigma)$ for
the absolutely continuous distribution of $\mY = \hvec^{-1}(\mZ)$ generated
by a latent multivariate normal variable $\mZ = \hvec(\mY) \sim \ND_J(\nullvec, \mSigma)$. 
\added{The copula parameters $\mSigma$ are identified as a consequence of the uniqueness of
the copula in Sklar's theorem.}
We allow more general sample spaces for binary, ordered, count, or otherwise discrete variables
and mixed continuous-discrete variables and thus neither require $\samY_j = \RR$ for
$j = 1, \dots, J$ nor the existence of $\hvec^{-1}$. 
\added{In this discrete situation, identifiability issues arise with two-step
copula estimators, however, maximum likelihood estimation has been demonstrated
to work in special cases \citep[Section 6.2.,][]{Genest_Neslehova_2007}.}

The model is invariant with respect to rescaling, that is
\begin{eqnarray*}
\text{NPN}(\hvec, \mSigma) = \text{NPN}\left(\diag(\dvec)^{-1} \hvec, \diag(\dvec) \mSigma \diag(\dvec)\right)
\end{eqnarray*}
for all $\dvec = (d_1, \dots, d_J)^\top \in \RR^J$ with $d_j > 0$ for all $j = 1, \dots, J$
and $\text{diag}(\dvec)$ the $J \times J$ diagonal matrix.
Thus, identifiability constraints on $\mSigma$ are needed. One option is to require
$\mSigma_{jj} \equiv 1$ for $j = 1, \dots, J$ leading to the interpretation
of $\h_j$ as probit-transformed marginal distribution function
$\Prob(\rY_j \le \ry_j) = \Phi(\h_j(\ry_j))$ for all $\ry_j \in \samY_j, j = 1, \dots,
J$.

Alternatively, we write $\mSigma = \mLz^{-1} \mLz^{-\top}$ in terms of the
inverse lower triangular Cholesky factor $\mLz^{-1}$ of the covariance matrix $\mSigma$
and require $\mLz_{jj} \equiv 1$ for $j = 1, \dots, J$.
This implies $\mSigma_{11} \equiv 1$ and $\mSigma_{jj} \ge 1$ for $j = 2, \dots, J$
and we define $\mPhi_\mLz := \mPhi_{\mLz^{-1} \mLz^{-\top}}$. 
In the absolutely continuous case with $\mY \in \RR^J$, the model 
$\mY \sim \text{NPN}(\hvec, \mLz^{-1} \mLz^{-\top})$ is identical to a structural equation model defined by a series 
of additive transformation models beginning with the marginal model 
$\Prob(\rY_1 \le \ry_1) = \Phi(\h_1(\ry_1))$ and proceeding
with conditional models
\begin{eqnarray*}
& & \Prob(\rY_j \le \ry_j \mid \rY_1 = \ry_1, \dots, \rY_{j-1} = \ry_{j-1}) =
\Phi\left(\sum_{\jmath = 1}^j \mLz_{j\jmath} \h_\jmath(\ry_\jmath)\right),
\quad j = 2, \dots, J
\end{eqnarray*}
for any $(\ry_1, \dots, \ry_J)^\top \in \RR^J$. \added{In other words, 
the rows of $\mLz$ contain regression coefficients
for the conditional distribution of $\rY_j$ given its predecessors and
the transformation $\h_j(\rY_j)$ is \emph{conditionally} standard normal.}
For exclusively binary outcomes $\samY_j = \{0, 1\}, j = 1, \dots, J$ we have
$\Prob(\rY_j = 0 \, \forall j = 1, \dots, J) = \mPhi_\mSigma(\parm)$,
where $\parm = (\eparm_1, \dots, \eparm_J)^\top \in \RR^J$ with $\eparm_j =
\sqrt{\mSigma_{jj}}\Phi^{-1}(\Prob(\rY_j = 0))$.

In the presence of covariates $\rX = \rx \in \samX$, one can characterise
the model via the conditional joint cumulative distribution function
\begin{eqnarray} \label{fm:xcdf}
\Prob(\mY \le \yvec \mid \rX = \rx) = \mPhi_{\mSigma(\rx)}(\hvec(\yvec \mid \rx))
\end{eqnarray}
where the covariates impact the transformation functions $\hvec(\yvec \mid
\rx)$, the covariance matrix $\mSigma(\rx)$, or both. For example, simple probit
linear transformation models
\begin{eqnarray} \label{fm:BoxCox}
\h_j(\ry_j \mid \rx) = \h_j(\ry_j) - \rx^\top \shiftparm_j, \quad j \in \{1, \dots, J\}
\end{eqnarray}
feature linear covariate effects. More elaborate choices include
transformation functions $\h_j(\ry_j \mid \rx)$ of the form $ \Phi^{-1}(F_j(\h_j(\ry_j) - \rx^\top
\shiftparm_j))$,
where $F_j: \RR \rightarrow [0,1]$ denotes an absolutely continuous
distribution function with log-concave density. For example, a marginal Weibull model can be
formulated via the inverse complementary log-log ($F_j = \text{cloglog}^{-1}$)
and a log-linear function $\h_j(\ry_j)$ \citep[see Table~1 in][]{Hothorn_Moest_Buehlmann_2017}.
Also the joint distribution might change with $\rx$, for example via linear models
for the off-diagonal elements of the inverse Cholesky factor
\begin{eqnarray} \label{fm:Lx}
\mLz_{j\jmath}(\rx) = \left\{ \begin{array}{ll}
  1 & 1 \le j = \jmath \le  J\\
  \alpha_{j\jmath} + \rx^\top \gammavec_{j\jmath} & 1 \le \jmath < j \le J.
\end{array} \right.
\end{eqnarray}
In the context of multivariate transformation models, such a
parameterisation has been proposed by \cite{Klein_Hothorn_Barbanti_2020}. 
For multivariate normal distributions, the same idea was applied by \cite{Barrathh_Boyd_2023}. 
For the sake of notational simplicity, we will consider the unconditional
case in Sections~\ref{sec:param} to \ref{sec:empeval} and comment on such conditional extensions
in \applications.

\section{Parameterisation} \label{sec:param}

The term ``nonparanormal'' insinuates a combination of nonparametrically
parameterised marginal distributions with a parametric Gaussian copula. 
As a gold standard, we therefore first derive the ``nonparanormal''
log-likelihood via nonparametric margins from $N$ independent samples $\mY_1, \dots, \mY_N \sim \text{NPN}(\hvec, \mSigma)$
with realisations $\mY_i = (\rY_{i1}, \dots, \rY_{iJ})^\top$ whose elements
were drawn from the
corresponding sample space $\rY_{ij} \in \samY_j, i = 1, \dots, N; j = 1,
\dots, J$.

In the model $\text{NPN}(\hvec, \mSigma)$, we first parameterise the $j = 1, \dots, J$ transformation functions
$\h_j$. In each dimension $j = 1, \dots, J$, we consider the ``empirical'' sample
space given by the ordered unique realisations $\upsilonvec_j = \{\upsilon_{j1}, \dots,
\upsilon_{jK(j)}\} \subseteq \samY_j$ with $\upsilon_{j,k - 1} < \upsilon_{j,k}$ for $k =
2, \dots, K(j)$. For the $i$th observation in the $j$th variable, write
$\rk(i,j) \in \{1, \dots, K(j)\}$ such that $\rY_{ij} = \upsilon_{j,\rk(i,j)}$.
In the absence of ties, $\rk(i,j)$ is the rank of the $i$th observation in
the sample $\rY_{1j}, \dots, \rY_{Nj}$. We can now parameterise the
transformation function $\h_j$ as a step function $\h_j(\upsilon_{jk}) = \eparm_{jk} \in \RR$ for $k = 0, \dots, K(j)$ with
values $\eparm_{j0} \equiv -\infty$ and $\eparm_{jK(j)} \equiv \infty$ at
the boundaries. The $j$th marginal parameter vector $\parm_j =
(\eparm_{j1}, \dots, \eparm_{j,K(j) - 1})^\top \in \RR^{K(j) - 1}$ comes
with a monotonicity constraint $\mD_j \parm_j \ge \nullvec_{K(j) - 2}$
defined by the $(K(j) - 2) \times (K(j) - 1)$ first order difference matrix $\mD_j$.
Finally, we collect all marginal parameters in the parameter vector $\parm =
(\parm_1^\top, \dots, \parm_J^\top)^\top \in \RR^{\sum_{j = 1}^{J} (K(j) -
1)}$ fully specifying $\hvec$. \added{Because the number of parameters $K(j)$ grows
with the sample size whenever the sample space is infinite, the
situation here is typically referred to as being ``nonparametric'',
although the vector $\parm$ contains parameters, in fact, potentially many of
them. The connection to the ``parametric'' situation will be made in
the next section.}

Second, we parameterise the inverse Cholesky factor $\mLz$ of the 
covariance matrix $\mSigma = \mLz^{-1} \mLz^{-\top}$ by defining
a lower triangular unit matrix $\mLambda = \mLambda(\lambdavec)$
in terms of its unconstrained lower triangular elements
$\lambdavec = (\lambda_{21}, \lambda_{31}, \dots, \lambda_{J,J-1})^\top \in \RR^{\nicefrac{J(J-1)}{2}}$. 
The first option to ensure parameter identifiability is to write 
$\mLz = \mLz^{(1)}(\lambdavec) = \mLambda$ to obtain $\mLz_{jj} \equiv 1$
for $j = 1, \dots, J$ \added{was already mentioned}. As a second option, we can write $\mLz = \mLz^{(2)}(\lambdavec) = \mLambda
\diag(\mLambda^{-1} \mLambda^{-\top})^{\nicefrac{1}{2}}$ such that
$\mSigma = \mLz^{-1} \mLz^{-\top}$ is equal to $\diag(\mLambda^{-1} \mLambda^{-\top})^{-\nicefrac{1}{2}} \mLambda^{-1} \mLambda^{-\top} \diag(\mLambda^{-1}
\mLambda^{-\top})^{-\nicefrac{1}{2}}$ ensuring the condition $\mSigma_{jj} \equiv 1, j = 1, \dots, J$. 
\added{Consequently, $\h_j(\rY_j)$ is now \emph{marginally} standard normal in the
absolute continuous case. The transformation to conditional standard
normality arises from $\mLz^{(2)}_{jj} \h_j(\rY_j)$.}
In the following we refer to these two options as $\mLz^{(s)}$ for $s = 1, 2$.
In either case, $\mSigma$ is parameterised in terms of the $\nicefrac{J(J - 1)}{2}$ lower
triangular parameters $\lambdavec$ of $\mLambda$ and is, for all values of
$\lambdavec$, symmetric and positive semidefinite. \added{Furthermore,
if $\lambda_{j\jmath} = 0$, then $\rY_j$ and $\rY_\jmath$ are conditionally
independent given the remaining variables in $\rY_1, \dots, \rY_j$ in both
parameterisations \citep[the reverse might not be true when some margins are
discrete, see][Section 4.1.]{Genest_Neslehova_2007}.}

\section{Nonparanormal Log-likelihoods} \label{sec:ll}

\added{In the following, we use the term ``likelihood'' in its broadest
sense, including Fishers original \emph{definition} of it as being
proportional to the probability of the observed data and
\emph{approximations} thereof in terms of absolutely continuous densities
\citep[][Section 5]{Lindsey_1999}.  This point of view allows us to choose
between ``empirical'' or ``nonparametric'' and ``parametric'' variants (or
combinations thereof) motivated by practical necessity and computational convenience.}

Before deriving the joint log-likelihood for all $J$ variables, we consider the
likelihood for $\parm_j$, that is, the parameters defining the $j$th marginal distribution. For the absolutely
continuous case recall that the empirical or nonparametric log-likelihood
\citep{Owen_2001} given by
\begin{eqnarray} \label{fm:emploglik}
\ell_j(\parm_j) = \sum_{i = 1}^N \log\left(\Phi(\eparm_{j,\rk(i,j)}) - \Phi(\eparm_{j,\rk(i,j) - 1})\right)
= \sum_{i = 1}^N \log\left(\,\int\limits_{\eparm_{j,\rk(i,j) - 1}}^{\eparm_{j,\rk(i,j)}} \phi(\rz) \, d\rz \right)
\end{eqnarray}
leads to a convex problem whose analytical solution $\Phi(\hat{\eparm}_{j,\rk(i,j)}) = \rk(i,j) / N$
is identical to the empirical cumulative distribution function evaluated that
$\upsilon_{j,r(i,j)}$. \added{From this perspective, many ``nonparametric''
estimators can be motivated as maximum \emph{empirical} likelihood
estimators and we will follow this path with the nonparanormal
log-likelihood to be introduced in Formula~(\ref{fm:NPL}).}

Furthermore, assume we had directly observed the latent
multivariate normal variables $\mZ_i \sim \ND_J(\nullvec, \mLz^{-1} \mLz^{-\top})$ 
with absolute continuous density
\begin{eqnarray} \label{fm:lognorm}
\phivec(\zvec \mid \mLz) & = & \exp\left(-\frac{J}{2}\log(2\pi) + 
                                  \sum_{j = 1}^J \log(\mLz_{jj}) 
                                  -\frac{1}{2} \lVert \mLz \zvec \rVert_2^2 \right) \\
& & \diag(\mLz) > \nullvec, \mLz \in \RR^{J \times J} \, \text{lower triangular} \nonumber
\end{eqnarray}
for $i = 1, \dots, N$. Then, the parametric negative log-likelihood 
$-\tilde{\ell}_J^{(0)}(\mLz) = -\sum_{i = 1}^N \tilde{\ell}_{J,i}^{(0)}(\mLz)$ with
\begin{eqnarray*} 
\tilde{\ell}_{J,i}^{(0)}(\mLz) & = & \log(\phivec(\mZ_i \mid \mLz)) \propto -\frac{1}{2} \lVert \mLz
\mZ_i \rVert_2^2 + \sum_{j = 1}^J \log(\mLz_{jj}) 
\end{eqnarray*}
is convex in $\mLz$ \citep{Barrathh_Boyd_2023}. In this section, we leverage
both principles to define a novel log-likelihood for the NPN model.

\paragraph{The Nonparanormal Log-likelihood.}

The nonparanormal log-likelihood for all $J$ variables is a direct extension
of the bivariate log-likelihood for ordinal data proposed by
\cite{Joergeskog_1994}. By replacing the univariate standard normal
density $\phi$ in the nonparametric log-likelihood~(\ref{fm:emploglik}) with the $J$-dimensional density 
$\phivec(\zvec \mid \mLz)$ of $\ND_J(\nullvec, \mLz^{-1} \mLz^{-\top})$ while keeping the integration limits for the $j$th
dimension in a $J$-dimensional integral, we define the nonparanormal log-likelihood by
$\ell_J^{(s)}(\parm, \lambdavec) = \sum_{i = 1}^N \ell^{(s)}_{J,i}(\parm,
\lambdavec)$ with
\begin{eqnarray} \label{fm:NPL}
\ell^{(s)}_{J,i}(\parm, \lambdavec) 
 & = & \log\left(\Prob_{\parm, \lambdavec}\left(\bigcap_{j = 1}^J
\eparm_{j,\rk(i,j) - 1} < \h_j(\rY_{ij}) \le \eparm_{j,\rk(i,j)} \right) \right) \\ \nonumber
 & = & \log\left(\Prob_{\mLz^{(s)}(\lambdavec)}\left(\hvec(\mY_i) \in \pbox_i(\parm)\right)\right) \\ \nonumber
 & = & \log\left(\int_{\pbox_i(\parm)} \phivec\left(\zvec \mid \mLz^{(s)}(\lambdavec)\right) \, d
\zvec\right), \quad s \in \{1, 2\} \nonumber
\end{eqnarray}
where $\pbox_i(\parm) = \{\zvec \in \RR^J \mid \eparm_{j,\rk(i,j) - 1} < \rz_j \le
\eparm_{j,\rk(i,j)}; j = 1, \dots, J\}$. We refer to this nonparanormal log-likelihood $\ell_J^{(s)}(\parm,
\lambdavec)$ as ``\NPN''.

\paragraph{The Smooth Nonparanormal Log-likelihood.}

The number of parameters $K(j)$ grows with $N$ for infinite sample spaces
$\samY_j$ and one might want to reduce the number of parameters in such cases. 
For $\samY_j \subseteq \RR$, define $\eparm_{jk} = \h_j(\upsilon_{jk} \mid \varparm_j) = \basisy_j(\upsilon_{jk})^\top
\varparm_j$ in terms of a spline basis $\basisy_j: \samY_j \rightarrow \RR^{\dimparm(j)}$ and
corresponding coefficients $\varparm_j \in \RR^{\dimparm(j)}$, potentially under some constraint $\mD_j
\varparm_j \ge \nullvec$. Typically, $\dimparm(j) < K(j)$. For finite discrete
sample spaces $\samY_j$, we use the same notation with $\evarparm_{jk} = \eparm_{jk}$ and $\h_j(\upsilon_{jk}
\mid \varparm_j) = \evec_{K(i)}(k)^\top \varparm_j$, where
$\basisy_j(\upsilon_{jk}) = \evec_{K(i)}(k)$ denotes the unit vector of
length $K(j)$ with non-zero element $k$ and $\dimparm(j) = K(j)$. \added{For
a fixed sample size, the empirical sample space discussed in
Section~\ref{sec:param} is finite discrete and we can thus establish a one-to-one
connection between discrete and nonparametrically parameterised continuous
models.} Motivations for and examples of such parameterisations can be found in 
\cite{Hothorn_Moest_Buehlmann_2017}.
Let $\parm_j = \parm_j(\varparm_j)$ and
$\parm(\varparm) = (\parm_1(\varparm_1)^\top, \dots,
\parm_J(\varparm_J)^\top)^\top$
for $\varparm = (\varparm_1, \dots, \varparm_J)^\top$ and define the
log-likelihood $\ell_J^{(s)}(\varparm, \lambdavec) := \ell_J^{(s)}(\parm(\varparm), \lambdavec)$.
Because the bases $\basisy_j$ and thus the transformations
$\basisy_j(\ry_j)^\top \varparm_j$
are smooth in $\ry_j$, we refer to the log-likelihood $\ell_J^{(s)}(\varparm,
\lambdavec)$ as ``\sNPN'', \added{allowing us to estimate a fully parametric model
featuring a fixed number of parameters by maximising an empirical likelihood.}

\paragraph{The Flow Nonparanormal Log-likelihood.}

When all response variables are absolutely continuous, we can approximate
the \sNPN defined through log\hyp{}probabilities by the corresponding multivariate
log-densities. The density in the distribution function
\begin{eqnarray*}
\Prob(\mY \le \yvec) & = & \mPhi_{\mLz}(\hvec(\yvec)) 
  = \int\limits_{-\infty}^{\hvec(\yvec)} \phivec(\zvec \mid \mLz) \, d \zvec
= \int\limits_{-\infty}^\yvec \phivec(\hvec(\tilde{\yvec}) \mid \mLz)
\text{det}(\hvec^\prime(\tilde{\yvec}))\, d \tilde{\yvec}
\end{eqnarray*}
motivates the approximate log-likelihood 
\begin{eqnarray*}
\tilde{\ell}_J^{(s)}(\varparm, \lambdavec) & = & \sum_{i = 1}^N
\log\left(\phivec\left((\h_1(\rY_{i1} \mid \varparm_1), \dots,
\h_J(\rY_{iJ} \mid \varparm_J))^\top \mid \mLz^{(s)}(\lambdavec)\right)\right) +  \\
& & \quad \sum_{j = 1}^J \log(\h_j^\prime(\rY_{ij} \mid \varparm_j)), \quad s \in \{1, 2\}
\end{eqnarray*}
where $\h_j(\rY_{ij} \mid \varparm_j) = \basisy_j(\rY_{ij})^\top \varparm_j$
and $\h_j^\prime(\rY_{ij} \mid \varparm_j) = \basisy^\prime_j(\rY_{ij})^\top
\varparm_j$ \citep[][]{Hothorn_Moest_Buehlmann_2017}. Because $\mLz \hvec(\mY) \sim \ND_J(\nullvec, \mI)$ is a simple
normalising flow \citep{JMLR:v22:19-1028}, we use the term ``\fNPN'' for $\tilde{\ell}_J^{(s)}(\varparm,
\lambdavec)$.

\paragraph{The Mixed Nonparanormal Log-likelihood.}

If some response variables are discrete and some absolutely continuous, one can approximate
the absolutely continuous parts by the corresponding \fNPN in a mixed continuous-discrete
log-likelihood.
Without loss of generality, assume that the first $1 \le \jmath < J$
variables $\rY_\jmath$ are absolutely continuous and the remaining $J - \jmath$
variables are discrete. We first partition the inverse Cholesky factor 
\begin{eqnarray*}
\mLz = \begin{pmatrix}
\mA & \nullvec \\
\mB & \mC
\end{pmatrix}
\end{eqnarray*}
with the continuous $\mA \in \RR^{\jmath \times \jmath}$ and discrete $\mC \in \RR^{(J - \jmath) \times
(J - \jmath)}$ parts being lower triangular and the full
matrix $\mB \in \RR^{(J - \jmath) \times \jmath}$ representing the interplay
between continuous and discrete variables.  We then obtain $(\rY_1, \dots, \rY_\jmath)^\top \sim \text{NPN}((\h_1, \dots,
\h_\jmath)^\top, \mA^{-1} \mA^{-\top})$, a NPN model for the
continuous part with \fNPN $\tilde{\ell}_{\jmath}^{(s)}(\varparm_1, \dots, \varparm_\jmath, \lambda_{21}, \dots, \lambda_{\jmath,\jmath -1})$, and 
the conditional distribution of discrete given continuous variables
\begin{eqnarray*}
\rY_{\jmath + 1}, \dots, \rY_J \mid \rY_1 = \ry_1, \dots, \rY_\jmath = \ry_\jmath \sim \text{NPN}\left((\h_{\jmath + 1},
\dots, \h_J)^\top - \muvec, \mC^{-1} \mC^{-\top}\right)
\end{eqnarray*}
with $\muvec =  -\mC^{-1} \mB (\h_1(\ry_1), \dots, \h_\jmath(\ry_\jmath))^\top \in \RR^{J -
\jmath}$, that is, a NPN model for the discrete part given the realisations
of the continuous variables.
The log-likelihood contribution of all variables is then the sum of
$\tilde{\ell}_{\jmath,i}^{(s)}(\varparm_1, \dots, \varparm_\jmath, \lambda_{21}, \dots, \lambda_{\jmath,\jmath -1})$
and the term
\begin{eqnarray*}
& & \ell_{J,i\mid\jmath}^{(s)}(\varparm_1, \dots, \varparm_\jmath, \parm_{\jmath + 1}, \dots, \parm_J, \lambda_{\jmath +1, 1}, \dots, \lambda_{J,J-1}) = \\
& & \quad \log\left(\int_{\pbox_i(\parm_{\jmath + 1}, \dots, \parm_J)} \phivec\left(\zvec -
\muvec \mid \mC^{(s)}\right) \, d \zvec\right), \quad s \in \{1, 2\}
\end{eqnarray*}
where $\pbox_i(\parm_{\jmath + 1}, \dots, \parm_J) = \{\zvec \in \RR^{J-\jmath} \mid \eparm_{j,\rk(i,j) - 1} < \rz_j \le
\eparm_{j,\rk(i,j)}; j = \jmath + 1, \dots, J\}$. Here,
$\muvec = \muvec(\varparm_1, \dots, \varparm_\jmath, \lambda_{\jmath +1, 1}, \dots,
\lambda_{J,J-1}) = -\mC^{{(s)}^{-1}} \mB^{(s)} (\h_1(\ry_1 \mid \varparm_1), \dots, \h_\jmath(\ry_\jmath \mid
\varparm_\jmath))^\top \in \RR^{J - \jmath}$ depends on $\mB^{(s)}$ and
$\mC^{(s)}$ which, in turn, depend on $(\lambda_{\jmath +1, 1}, \dots, \lambda_{J,J-1})$. In total, we have
\begin{eqnarray*}
\tilde{\ell}^{(s)}_{J\mid\jmath}(\varparm_1,
\dots, \varparm_\jmath, \parm_{\jmath}, \dots, \parm_J, \lambdavec) & = &
\sum_{i = 1}^N \tilde{\ell}_{\jmath,i}^{(s)}(\varparm_1, \dots, \varparm_\jmath, \lambda_{21}, \dots, \lambda_{\jmath,\jmath -1}) +
\\
& & \quad \ell_{J,i\mid\jmath}^{(s)}(\varparm_1, \dots, \varparm_\jmath, \parm_{\jmath + 1}, \dots, \parm_J, \lambda_{\jmath +1, 1}, \dots, \lambda_{J,J-1})
\end{eqnarray*}
and we refer to this form of the log-likelihood as ``\mNPN''.

In summary, we defined the \NPN $\ell_J^{(s)}(\parm,
\lambdavec)$ and three approximations thereof.  The \sNPN
$\ell_J^{(s)}(\parm(\varparm), \lambdavec)$ computes
log-probabilities based on smooth transformations, the \fNPN 
$\tilde{\ell}_J^{(s)}(\varparm, \lambdavec)$ for absolute continuous responses approximates log-probabilities
by log-densities, and the \mNPN, given by the term 
$\tilde{\ell}^{(s)}_{J\mid\jmath}(\varparm_1,
\dots, \varparm_\jmath, \parm_{\jmath}, \dots, \parm_J, \lambdavec)$, mixes
the \fNPN defined by log-densities of the
marginal distribution for $j = 1, \dots, \jmath$ with log-probabilities for
the remaining elements (which again can be in form of a \NPN or a \sNPN).  Each of these log-likelihoods
can be coupled with either constraint $s = 1$ (unit diagonal in $\mLz$) or
$s = 2$ ($\mSigma$ being a correlation matrix). Ways to enhance these log-likelihoods to covariate effects
in $\hvec$ or $\mLz$ are discussed in \applications.

\paragraph{Evaluation of Log-likelihood and Score Functions.}

Computing the \fNPN involves simple matrix multiplications whose
gradient with respect to $\varparm$ is
\begin{eqnarray*}
\frac{\partial \tilde{\ell}_{J,i}^{(s)}(\varparm, \lambdavec)}{\partial \varparm_j} & = & 
- \frac{1}{2} \mLz^{(s)}(\lambdavec)^\top \mLz^{(s)}(\lambdavec) 
  (\basisy_1(\rY_{i1})^\top \varparm_1, \dots, \basisy_J(\rY_{iJ})^\top \varparm_J)^\top \basisy_j(\rY_{ij})^\top +
  \frac{\basisy^\prime(\rY_{ij})^\top}{\basisy^\prime_j(\rY_{ij})^\top \varparm_j}
\end{eqnarray*} 
both for $s = 1$ and $s = 2$. For $s = 1$, the score
function with respect to $\lambdavec$ is 
\begin{eqnarray*}
\frac{\partial \tilde{\ell}_{J,i}^{(1)}(\varparm, \lambdavec)}{\partial
\lambda_{j\jmath}} & = & 
- \left(\mLambda (\basisy_1(\rY_{i1})^\top \varparm_1, \dots,
\basisy_J(\rY_{iJ})^\top) (\basisy_1(\rY_{i1})^\top \varparm_1, \dots,
\basisy_J(\rY_{iJ})^\top \varparm_J)^\top \right)_{j\jmath}
\end{eqnarray*}
for $1 < \jmath < j < J$. The case of $s = 2$ is more elaborate, the score
function is derived in \cite{vign:mvtnorm} listed in the supplementary
material, see Section~\ref{app:impl}.

The \NPN, \sNPN, and \mNPN require the evaluation
of multivariate normal probabilities over boxes $\pbox_i$ and the algorithm by
\cite{Genz_1992} has been widely applied to approximate such probabilities
by quasi-Monte-Carlo integration. The only attempt to also approximate the score function
for these log-probabilities we are aware of was described for the special case of
binary outcomes by \cite{Christoffersen_2021}. \deleted{Instead of approximating both
the log-likelihood and the corresponding score function, we propose to approximate the
log-likelihood by Genz' method in a first step and, in a second step, 
to derive the exact score function of this approximation rather than an
approximate score function of the true log-likelihood.}

In this simplest form, for $\pbox_i = \{\zvec \in \RR^J \mid \ubar{\bvec} < \zvec \le
\bar{\bvec}\}$, the probability defining the \NPN contribution is approximated as
\begin{eqnarray*}
\exp\left(\ell_{J,i}^{(1)}(\parm, \lambdavec)\right) & \approx & \Ex_\mW \prod_{j = 1}^J (e_j(\mW) - d_j(\mW)) \quad \text{with} \\
d_j(\mW) & = & \Phi_1\left(\ubar{b}_j - \sum_{\jmath = 1}^{j-1} \mLambda^{-1}_{\jmath j}\Phi^{-1}(d_\jmath +
W_\jmath(e_\jmath(\mW) -
d_\jmath(\mW)))\right); \quad d_1(\mW) = \Phi(\ubar{b}_1)\\
e_j(\mW) & = & \Phi_1\left(\bar{b}_j - \sum_{\jmath = 1}^{j-1} \mLambda^{-1}_{\jmath j}\Phi^{-1}(d_\jmath +
W_\jmath(e_\jmath(\mW) - 
d_\jmath(\mW)))\right); \quad e_1(\mW) = \Phi(\bar{b}_1)
\end{eqnarray*}
and the expectation is over $\mW = (W_1, \dots, W_{J-1})^\top \in \RR^{J - 1}$,
$W_j \sim \UD(0,1)$ whose elements are independent. The expectation in turn is approximated by the mean
over $M$ independent draws of $\mW$. \added{The question arises how many samples
$M$ are required when searching for the maximiser of the sample log-likelihood
\begin{eqnarray*}
N^{-1} \sum_{i = 1}^N \ell_{J,i}^{(1)}(\parm, \lambdavec).
\end{eqnarray*}
Conceptually, because $\mY_1, \dots, \mY_N$ were assumed to be independent, the variance
of this mean of log-probabilities $\ell_{J,i}^{(1)}(\parm, \lambdavec)$ 
converges to zero as the number of observations $N$ increases, regardless of
$M$. For finite samples, of course, one should assess the impact of $M$ on
the maximum likelihood estimators for $\parm$ and $\lambdavec$ and choose
$M$ large enough. \cite{Graz_Barbanti_Brandl_2026} reports extensive simulation results
indicating that, depending on context, maximum likelihood estimation
may being to stabilise with values as small as $M = 250$ or $M = 500$.}

\added{Instead of approximating both
the log-likelihood and the corresponding score function, we propose to approximate the
log-likelihood by Genz' method in a first step and, in a second step, 
to derive the exact score function of this approximation rather than an
approximate score function of the true log-likelihood.}
\replaced{Conditionally on $M$}{For given} realisations of $\mW$, the score function
with respect to $\ubar{\bvec}$ and $\bar{\bvec}$ and the score function 
with respect to the lower off-diagonal elements of $\mLambda^{-1}$ can
be computed \added{deterministically}
by the chain-rule. Scores with respect to $\mLambda$ are
then given by $-\mLambda^{-\top} \otimes \mLambda^{-1}$.
A modular re-implementation of \cite{Genz_1992} algorithm 
and its score function, also for $s = 2$, is given in \cite{vign:mvtnorm} as
listed in Section~\ref{app:impl}. 

\section{Properties and Convex Approximations} \label{sec:convex}

Unfortunately, neither of these nonparanormal log-likelihoods leads to a convex
optimisation problem. We study the properties of each of the four
log-likelihoods in the following theorems. These disappointing results,
however, lead to some insights allowing to suggest some convex
approximations to these problems.


We first consider the \fNPN $\tilde{\ell}_J^{(s)}(\varparm, \lambdavec)$,
which only involves multivariate normal densities, transformation
functions, and derivatives thereof.

\begin{thm} \label{thm:fNPN}
Minimising $ -\tilde{\ell}_J^{(s)}(\varparm, \lambdavec)$ 
subject to $\mD_j \varparm_j \ge \nullvec$ for $j = 1, \dots, J$
is a biconvex problem in $\varparm \in \RR^{\sum_{j = 1}^{J} (P(j) - 1)}$ and $\lambdavec
\in \RR^{\nicefrac{J(J - 1)}{2}}$ for $s = 1, 2$.
\end{thm}

The \added{negative} \NPN, defined by log-probabilities, is not necessarily convex in $\lambdavec$.

\begin{thm} \label{thm:NPN}
Minimising $-\ell_J^{(s)}(\parm, \lambdavec)$
subject to $\mD_j \parm_j \ge \nullvec$ for $j = 1, \dots, J$
is a convex problem in $\parm \in \RR^{\sum_{j = 1}^{J} (K(j) - 1)}$ for given $\lambdavec
\in \RR^{\nicefrac{J(J - 1)}{2}}$ for $s = 1, 2$. It is not necessarily
convex in $\lambdavec \in \RR^{\nicefrac{J(J - 1)}{2}}$.
\end{thm}

\added{Consequently, the \added{negative} \sNPN is also not necessarily convex in
$\lambdavec$.} Under independence (that is, for $\lambdavec = \nullvec$), the \NPN
$\ell_J^{(s)}(\parm, \nullvec) = \sum_{j = 1}^J \ell_j(\parm_j)$
is the sum of $J$ marginal empirical log-likelihoods and therefore $\parm_1,
\dots, \parm_J$ are orthogonal. This property is lost whenever $\lambdavec \neq \nullvec$
and therefore $\ell_J^{(s)}(\parm, \lambdavec)$ has to be maximised with respect to both $\parm$ and
$\lambdavec$ simultaneously for proper estimation and uncertainty
assessment. We conclude with a statement about the \mNPN.

\begin{coro} \label{coro:mNPN}
Minimising $-\tilde{\ell}^{(s)}_{J \mid \jmath}(\varparm_1, \dots, \varparm_\jmath, \parm_{\jmath}, \dots, \parm_J,
\lambdavec)$ subject to $\mD_j \varparm_j \ge \nullvec$ for $j = 1, \dots, J -
1$ and $\mD_J \parm_J \ge \nullvec$
is a biconvex problem for $s = 1$ and $\jmath = J - 1$, that is in $(\varparm_1, \dots,
\varparm_{J-1}, \parm_J) \in \RR^{K(J) - 1 + \sum_{j = 1}^{J - 1} (P(j) - 1)}$ and $\lambdavec
\in \RR^{\nicefrac{J(J - 1)}{2}}$ for $s = 1, 2$.
\end{coro}

The proofs are given in \SM~\ref{app:proofs}.


Given the malign nature of the optimisation problems involved, we discuss
three convex approximations which, at the very least, help to derive
good starting values:

\begin{enumerate}
\item \label{convex1} Minimise $-\ell_j(\parm_j)$,  or $-\ell_j(\parm_j(\varparm_j))$, 	in~(\ref{fm:emploglik}),
 and obtain the empirical marginal estimate $\hat{\parm}_j$ for all $j = 1, \dots, J$ and get
      $\hat{\parm} = (\hat{\parm}_1, \dots, \hat{\parm}_J)^\top$.
      One typically tries to avoid $\eparm_{j,N} = \infty$ by changing the
      estimator to the normal score $\rk(i,j) (N + 1)^{-1} = \Phi(\hat{\eparm}_{j,\rk(i,j)})$
      \citep[or a winsorised version,][]{Mai_He_Zou_2023}
      when estimating $\parm_j$ (this problem is not present when a smoothly
      parameterised model is given by $\hat{\varparm}_j$). Define 
$\hat{\mZ}_i = (\hat{\eparm}_{1,\rk(i,1)}, \dots,
\hat{\eparm}_{J,\rk(i,J)})^\top$ and 
minimise $-\tilde{\ell}_J^{(0)}(\mLz^{(2)}(\lambdavec))$
with respect to $\lambdavec$. This maintains the interpretation of
$\Phi(\hat{\eparm}_{j,\rk(i,j)})$ as $j$th marginal distribution function
evaluated at $\rY_{ij}$.

\item \label{convex2}
For the \fNPN, an iterative version with alternating estimation of
$\lambdavec$ or $\varparm$, that is, switching between the two target functions 
$\tilde{\ell}_J^{(s)}(\hat{\varparm}, \lambdavec)$ and
$\tilde{\ell}_J^{(s)}(\varparm, \hat{\lambdavec})$,
is known as alternate convex search (ACS) which, under certain
conditions, might converge \citep{Gorski_2007}.

\item \label{convex3}
\begin{enumerate} 
\item Solve the convex problem (in $\varparm_1$) and
minimise $-\tilde{\ell}^{(s)}_{1}(\varparm_1)$ subject to $\mD_1 \varparm_1 \ge \nullvec$.
\item Solve the convex problem (in $\varparm_2$ and $\lambda_{21}$) and
minimise $-\tilde{\ell}_2^{(s)}(\hat{\varparm}_1, \varparm_2, \lambda_{21})$
subject to $\mD_2 \varparm_2 \ge \nullvec$.
\item Solve the convex problem (in $\varparm_3$ and $\lambda_{3\cdot}$) and
minimise $-\tilde{\ell}_3^{(s)}(\hat{\varparm}_1, \hat{\varparm}_2,
\varparm_3, \hat{\lambda}_{21}, \lambda_{3\cdot})$ subject to $\mD_3 \varparm_3 \ge
\nullvec$.
\item Repeat until $j = J$. Solve the convex problem (in $\varparm_J$ and $\lambda_{J\cdot}$)
and minimise $-\tilde{\ell}_J^{(s)}(\hat{\varparm}_1, \hat{\varparm}_2, \dots,
\hat{\varparm}_{J-1}, \varparm_J, \hat{\lambda}_{21}, \dots, \hat{\lambda}_{(J - 1)\cdot},
\lambda_{J\cdot})$ subject to $\mD_J \varparm_J \ge \nullvec$.
\end{enumerate}
\end{enumerate}

The approaches in \ref{convex1}.~are variants of the two-step maximum pseudo
likelihood estimator, where the parameters of the marginal models are fitted
first. Treating these parameters as known, the copula parameters are 
then estimated in a second step. Two-step maximum pseudo likelihood
estimators are used as starting values for the simultaneous maximisation of
\NPN functions with respect to all model parameters in
Section~\ref{sec:empeval}.

The sequential approximation~\ref{convex3}.~fits a series of linear transformation 
models to regressors $\hat{\h}_1, \dots, \hat{\h}_{j - 1}$, where only $\varparm_j$ and the $j$th row
$\lambda_{j\cdot}$ of $\mLambda$ are updated.
This works for $s = 1, 2$ and also allows penalisation of the $\mLambda$
parameters for high(er)-dimensional data as suggested for normal models by \cite{Khare_2019}.
Variants~\ref{convex2}.~and~\ref{convex3}.~could also be combined with the smooth and \mNPN.

\section{Empirical Comparisons} \label{sec:empeval}

The theoretical and computational framework presented in
Sections~\ref{sec:model}--\ref{sec:convex} is too broad to be
empirically evaluated in an exhaustive way. We therefore focus on one
application \added{highlighting the versatility of the framework}
and a simple simulation setup to illustrate potential practical
merits. \added{A more comprehensive picture of the empirical performance of
maximum nonparanormal likelihood estimation emerges from simulation results
published elsewhere: \cite{Klein_Hothorn_Barbanti_2020} investigate the
\fNPN with covariate-dependent $\mLambda$, \cite{Graz_Barbanti_Brandl_2026} estimate and
evaluate multivariate count models obtained by maximisation of the \NPN, and the
\mNPN was evaluated for the estimation of time-dependent receiver 
operating characteristic curves \citep{Sewak_Inacio_Wuu_2026} and in 
covariate adjustment procedures \citep{Dandl_Hothorn_2026}.}

\subsection{Transformation Discriminant Analysis} \label{subsec:DA}

\begin{figure}
\begin{center}
\begin{knitrout}
\definecolor{shadecolor}{rgb}{0.969, 0.969, 0.969}\color{fgcolor}
\includegraphics[width=\maxwidth]{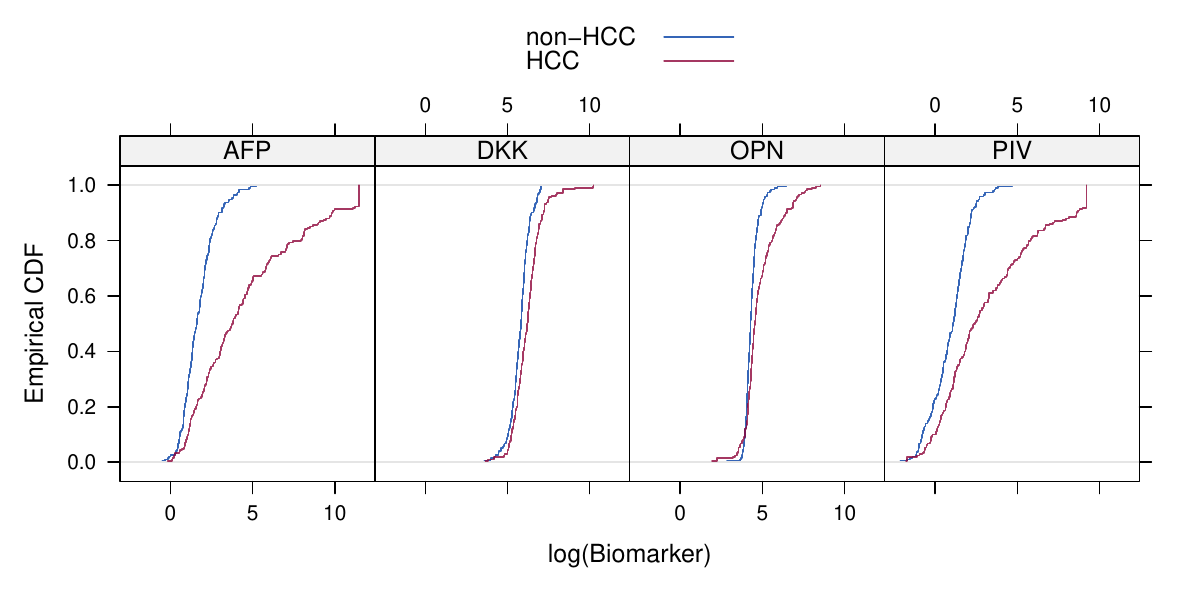} 
\end{knitrout}
\caption{Case-control study for Hepatocellular carcinoma (HCC) by
\cite{jang2016diagnostic}:  Empirical cumulative distribution functions
(CDFs) for
log-transformed biomarkers alpha-fetoprotein (AFP), protein induced by
vitamin~K absence or
antagonist-II (PIV), osteopontin (OPN), and Dickkopf-1 (DKK) in HCC cases
and non-HCC controls. 
\label{fig:HCC}}
\end{center}
\end{figure}

We discuss a discrimination function for hepatocellular carcinoma (HCC) diagnosis 
based on four biomarkers (DKK: Dickkopf-1, OPN:
osteopontin, PIV: protein induced by vitamin~K absence or
antagonist-II, and AFP: alpha-fetoprotein). Based on data reported 
from a retrospective case-control study of $403$ subjects by
\cite{jang2016diagnostic},
\cite{Sewak_Siegfried_Hothorn_2026} proposed the log-likelihood ratio function of a
transformation discriminant analysis (TDA) model as optimal discrimination function. The
empirical biomarker distributions presented for HCC cases and non-HCC
controls in Figure~\ref{fig:HCC} show that PIV and AFP readings are affected by a limit-of-detection
problem. For these subjects, it is only known that PIV (or AFP) is larger
than a specific detection limit, in other words, these observations are
right-censored.

\added{For the analysis of this dataset, we show how one can put together different models playing
with different parameterisations of the marginal and copula parameters and
then picking the `right' likelihood based on properties of the observations
(the \mNPN because of right-censoring).  The same case is made by the
\textsf{R} code snippet presented in Section~\ref{sec:comp}, demonstrating
how the theoretical concepts translate into software.}

We fit three models to the data. First, a classical linear discriminant
analysis (LDA) assuming a linear transformation function (and thus a linear basis
function $\basisy_j$) for each of the four
biomarkers and a common covariance, resulting in a joint normal distribution
of the biomarker values with class-specific means. Second, we replace
the linear transformation functions with potentially non-linear
ones~(\ref{fm:BoxCox})
featuring a location term differentiating between classes. As
a third option, we introduce a scale term such that the marginal variability
may differ between classes. In all models, we restrict our attention to a
common correlation matrix. 
With $x = 1$ for HCC case and $x = 0$ for a non-HCC control and $\mY =
(\rY_\text{AFP}, \rY_\text{DKK}, \rY_\text{OPN}, \rY_\text{PIV})^\top \in
\RR^4$ the LDA model is equivalent to 
$\mY \sim \ND_J(\mLz^{-1} \etavec(x), \mLz^{-1} \mLz^{-\top})$. Because this problem
is convex in both $\etavec(x)_j = \eshiftparm_j x$ and $\mLz$
\citep[Section~5.2.1.~in][]{Barrathh_Boyd_2023}, we use a convex solver
as a benchmark for later method comparison.

All NPN models feature variants of the transformation function
implementing a location-scale model \citep{Siegfried_Kook_Hothorn_2023} with
$\h_j(\ry_j \mid x) = \basisy_j(\ry_j)^\top \varparm_j \exp(\xi_j x) - \eshiftparm_{j} x$.
The LDA model can be formulated by choosing linear bases $\basisy_j^\top = (1, \ry_j)$ and
location-only part ($\xi_j = 0$ for all $j \in \{\text{DKK}, \text{OPN}, \text{PIV},
\text{AFP}\}$), however, this parameterisation leads to a non-convex
optimisation problem when minimising the
negative \fNPN $-\tilde{\ell}^{(2)}(\varparm, \lambdavec)$ simultaneously in
all model parameters $\varparm = (\varparm_\text{DKK}^\top,
\eshiftparm_\text{DKK}, \dots, \varparm_\text{AFP}^\top,
\eshiftparm_\text{APF})^\top$ and $\lambdavec$.

The log-likelihoods (normal negative convex and \fNPN) obtained by both optimisation routines are equivalent
($-2373.492$), this also applies to the log-likelihood ratios.
After dividing each column of $\hat{\mLz}$ obtained from the convex solver with the
square-root of the corresponding diagonal element, the estimated values of
$\lambdavec$ are also identical, see Table~\ref{tab:TDA}. This means that
the simultaneous one-step non-convex optimisation of the \fNPN resulted in the
same model as the optimisation of the corresponding convex negative normal
log-likelihood.

The location-only transformation discriminant analysis model (lTDA) is
obtained from more flexible basis functions; we use Bernstein polynomial
bases $\basisy_j^\top \in \RR^{7}$ of order $6$. 
The additional $20$ parameters introducing non-linear
transformations improve the \fNPN to 
$-2106.183$, at the expense of higher computing times (median $2.523$
instead of $0.706$ seconds). The maximum-likelihood estimates
$\hat{\lambdavec}$ are similar, as are the corresponding standard errors
obtained from the inverse Hessian.

A location-scale version of the above model introduces
four additional scale parameters $\xi_j$.
Again, an improvement in the \fNPN was observed ($-2069.909$), the computing time
increased only marginally (to median $3.143$ seconds).

The four models estimated via minimisation of the convex negative normal or \fNPN ignored the fact
that PIV or AFT biomarker values could not be observed for
$17$ subjects because the upper limit-of-detection was
reached.  Technically, these observations are right-censored, necessitating a
correction of the log-likelihood contributions for these observations.  We
implemented such a correction by a \mNPN combining the \fNPN for DKK and OPN
and a \sNPN with right-censoring for AFP and PIV, technically speaking,
the maximisation of 
\begin{eqnarray*}
\tilde{\ell}^{(2)}_{\text{PIV}, \text{AFP} \mid \text{DKK}, \text{OPN}}(
\varparm_\text{DKK}, \eshiftparm_\text{DKK}, \xi_\text{DKK}, \dots,
\parm_\text{AFP}(\varparm_\text{AFP}), \eshiftparm_\text{AFP}, \xi_\text{AFP},
\lambdavec).
\end{eqnarray*}
The in-sample 
\fNPN and \mNPN values are not directly comparable, however, 
neither the estimated
$\lambdavec$ parameters nor the corresponding standard errors are affected
by this more elaborate estimation, which took much longer to compute
(median $24.341$ seconds \added{with Halton sequences of length $M =
500$}). This result, obtained from the much
more elaborate \mNPN function, is practically important,
as it demonstrates that ignoring limits of detection \replaced{had}{has}
negligible influence on the fitted model in this application.


\begin{sidewaystable}
\begin{center}
\begin{tabular}{crrrrrrrrr} \hline \hline
& \multicolumn{3}{c}{LDA}  & \multicolumn{2}{c}{lTDA} & \multicolumn{4}{c}{lsTDA} \\
& convex  & \multicolumn{2}{c}{flow} & \multicolumn{2}{c}{flow}  & \multicolumn{2}{c}{flow} & \multicolumn{2}{c}{mixed} \\
$\lambda$ & $\hat{\lambda}$ & $\hat{\lambda}$ & $\text{SE}(\hat{\lambda})$ &$\hat{\lambda}$ & $\text{SE}(\hat{\lambda})$ &$\hat{\lambda}$ & $\text{SE}(\hat{\lambda})$ &$\hat{\lambda}$ & $\text{SE}(\hat{\lambda})$  \\ \hline
OPN,DKK  &  $ -0.180 $ &$ -0.180 $ &$ 0.051 $ &$ -0.107 $ &$ 0.050 $ &$ -0.099 $ &$ 0.050 $ &$ -0.099 $ &$ 0.050 $ \\ 
PIV,DKK  &  $ -0.330 $ &$ -0.330 $ &$ 0.053 $ &$ -0.296 $ &$ 0.053 $ &$ -0.300 $ &$ 0.053 $ &$ -0.300 $ &$ 0.053 $ \\ 
PIV,OPN  &  $ -0.319 $ &$ -0.319 $ &$ 0.052 $ &$ -0.344 $ &$ 0.053 $ &$ -0.311 $ &$ 0.053 $ &$ -0.311 $ &$ 0.053 $ \\ 
AFP,DKK  &  $ 0.019 $ &$ 0.019 $ &$ 0.053 $ &$ 0.032 $ &$ 0.053 $ &$ 0.041 $ &$ 0.053 $ &$ 0.041 $ &$ 0.053 $ \\ 
AFP,OPN  &  $ -0.083 $ &$ -0.083 $ &$ 0.053 $ &$ -0.171 $ &$ 0.054 $ &$ -0.186 $ &$ 0.054 $ &$ -0.186 $ &$ 0.054 $ \\ 
AFP,PIV  &  $ -1.945 $ &$ -1.945 $ &$ 0.109 $ &$ -1.428 $ &$ 0.099 $ &$ -1.345 $ &$ 0.097 $ &$ -1.345 $ &$ 0.097 $ \\ 

\hline
log-Lik & $-2373.492$ & \multicolumn{2}{r}{$-2373.492$} & \multicolumn{2}{r}{$-2106.183$} & \multicolumn{2}{r}{$-2069.909$} & \multicolumn{2}{r}{$-5268.329$} \\
time (sec) & $0.000$ & \multicolumn{2}{r}{$0.706$} & \multicolumn{2}{r}{$2.523$} & \multicolumn{2}{r}{$3.143$} & \multicolumn{2}{r}{$24.341$} \\
\hline \hline
\end{tabular}
\caption{HCC discriminant analysis: Linear discriminant analysis (LDA) and
transformation discriminant analysis (location-only: lTDA, location-scale:
lsTDA) fitted by minimisation of the convex negative log-likelihood of a
multivariate normal (for LDA only), \fNPN, and \mNPN, the latter taking
limit-of-detection problems into account. Maximum likelihood estimates for
$\lambdavec$ parameters and standard errors (via inverse observed Hessians)
are given. The in-sample log-likelihoods and median computing times are
reported in the bottom rows. \label{tab:TDA}}
\end{center}
\end{sidewaystable}

The \mNPN is not even biconvex and it might be interesting to look at the
results obtained by the convex approximations discussed in
Section~\ref{sec:convex}. The results in Table~\ref{tab:convex} suggest 
that alternating between the estimation of marginal and copula
parameters~(\ref{convex2}.)
provides a better approximation to the in-sample log-likelihood obtained by
simultaneous optimisation of all model parameters compared to the two-step
pseudo~(\ref{convex1}.) or sequential~(\ref{convex3}.) approaches. However,
the small mixed standard error $0.097$ of $\hat{\lambda}_{\text{AFP},\text{PIV}}$
(last row and column of Table~\ref{tab:TDA}) suggest that the discrepancies among estimates in rows of Table~\ref{tab:convex} might
be practically relevant. In addition, the alternation took twice as much
time as the simultaneous parameter estimation. Therefore, one might call
into question the practical merits of convex approximations in this setup.

\begin{table}[t!]
\begin{center}
\begin{tabular}{crrrr} \hline \hline
& \multicolumn{4}{c}{lsTDA}  \\
& simultaneous & pseudo~(\ref{convex1}.) & alternating~(\ref{convex2}.) & sequential~(\ref{convex3}.)  \\
$\lambda$ & $\hat{\lambda}$ & $\hat{\lambda}$ &$\hat{\lambda}$ &$\hat{\lambda}$ \\ \hline
OPN,DKK  &  $ -0.099 $ &$ -0.094 $ &$ -0.093 $ &$ -0.096 $ \\ 
PIV,DKK  &  $ -0.300 $ &$ -0.276 $ &$ -0.285 $ &$ -0.276 $ \\ 
PIV,OPN  &  $ -0.311 $ &$ -0.301 $ &$ -0.303 $ &$ -0.306 $ \\ 
AFP,DKK  &  $ 0.041 $ &$ 0.016 $ &$ 0.028 $ &$ 0.015 $ \\ 
AFP,OPN  &  $ -0.186 $ &$ -0.207 $ &$ -0.182 $ &$ -0.199 $ \\ 
AFP,PIV  &  $ -1.345 $ &$ -1.036 $ &$ -1.255 $ &$ -1.134 $ \\ 

\hline
log-Lik & $-5268.329$ & $-5295.388$ & $-5269.009$ & $-5282.360$ \\
time (sec) & $24.341$ & $0.702$ & $50.392$ & $10.890$ \\
\hline \hline
\end{tabular}
\caption{Location-scale transformation discriminant analysis (lsTDA) model for
HCC: simultaneous optimisation of marginal and copula parameters using the
\mNPN with three approximations: two-step pseudo, alternating, and sequential
(Section~\ref{sec:convex}). \label{tab:convex}}
\end{center}
\end{table}

\subsection{Polychoric Correlations} \label{subsec:poly}

%
%

For bivariate Gaussian copulas, the semiparametric efficiency bound is known
and the performance of several estimators against this
theoretical benchmark is studied in this section.
We sample $N \in \{10, 20, 50\}$ observations from $\mY = (\rY_1, \rY_2) \sim
\text{NPN}(\hvec, \mSigma(\rho))$ with $\h_1(\ry_1) = \Phi^{-1}(\chi^2_2(\ry_1))$ and $\h_2 =
\h_1$, that is, $\rY_j \sim \chi^2_2$ for $j = 1, 2$. The latent correlation 
between both variables is given by
$\mSigma(\rho) = ((1, \rho)^\top \mid (\rho, 1)^\top)$ for $\rho \in \{0, 0.1, 0.2, \dots,
0.9\}$.
\cite{Klaassen_Wellner_1997} established the semiparametric efficiency bound $(1 - \rho^2) /
\sqrt{N}$ 
for the correlation, that is, the standard error of semiparametric efficient
estimators $\hat{\rho}$. \added{It is identical to the standard error of the
Pearson correlation coefficient computed for the \emph{transformed}
bivariate observations $(\Phi^{-1}(\chi^2_2(\rY_{1i})),
\Phi^{-1}(\chi^2_2(\rY_{2i}))), i = 1, \dots, N$, that is, in the oracle case
where the marginal $\chi^2_2$ distribution functions are known.}

For both responses being absolutely continuous, we
estimate $\rho$ by the two-step maximum pseudo likelihood estimator \citep[which is,
according to][semiparametric efficient in this simple
case]{Klaassen_Wellner_1997}, with two different ways to obtain standard
errors (labelled ``ML'' and ``MPL'', see Section~\ref{sec:comp}).

We compare the performance of this estimator to maximum
likelihood-based estimation of $\rho$ using the \NPN, \sNPN, and
\fNPN, that is, by employing the transformation $\hat{\rho} = \nicefrac{-\hat{\lambda}_{21}}{\sqrt{1 +
\hat{\lambda}_{21}^2}}$. In addition, we also report estimators of the corresponding
standard errors of $\hat{\rho}$, obtained via the $\Delta$-method for the
procedures described in this paper.

To study the performance for non-continuous data, we transform each variable
to binary and ordinal (five categories) measurements using random empirical
quantiles between $20\%$ and $80\%$ as cut-offs.  As a competitor for binary
or ordinal variables, we use the composite likelihood
\citep[``cML'',][\added{which in the binary case is identical to the \NPN}]{Nikoloulopoulos_2023}.
In absence of a direct competitor for
mixed continuous-discrete responses, we only report the results obtained via
the \mNPN (that is, a mix of \NPN, \sNPN, or \fNPN for the continuous
variable and \NPN for the categorical variable).  For each combination of
measurement scales, we repeat the simulation $100$ times.

\begin{figure}
\begin{center}
\begin{knitrout}
\definecolor{shadecolor}{rgb}{0.969, 0.969, 0.969}\color{fgcolor}
\includegraphics[width=.8\linewidth]{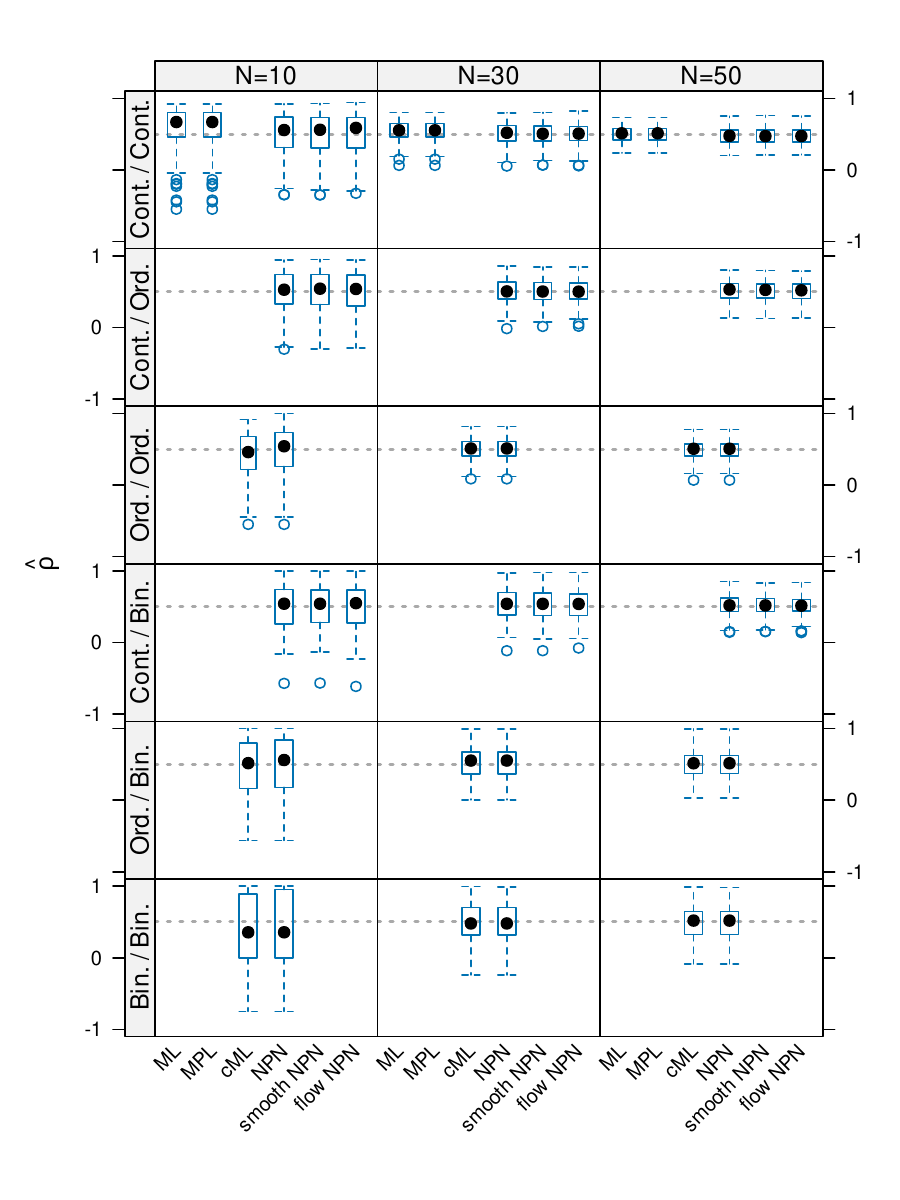} 
\end{knitrout}
\end{center}
\caption{Bivariate polychoric correlations: Distribution of $100$ estimators of $\rho = 0.5$
(horizontal line) from $N$ bivariate observations measured at different scales: continuous (Cont.), ordinal
(five levels, Ord.), and binary (Bin.). 
For continuous variables, the two-step maximum pseudo likelihood approach
(``ML'' and ``MPL'' provide identical estimators, but different standard errors)
is shown as competitor, for categorical variables, the composite maximum likelihood (cML)
is presented. \label{fig:rho}}
\end{figure}

\begin{figure}
\begin{center}
\begin{knitrout}
\definecolor{shadecolor}{rgb}{0.969, 0.969, 0.969}\color{fgcolor}
\includegraphics[width=.8\linewidth]{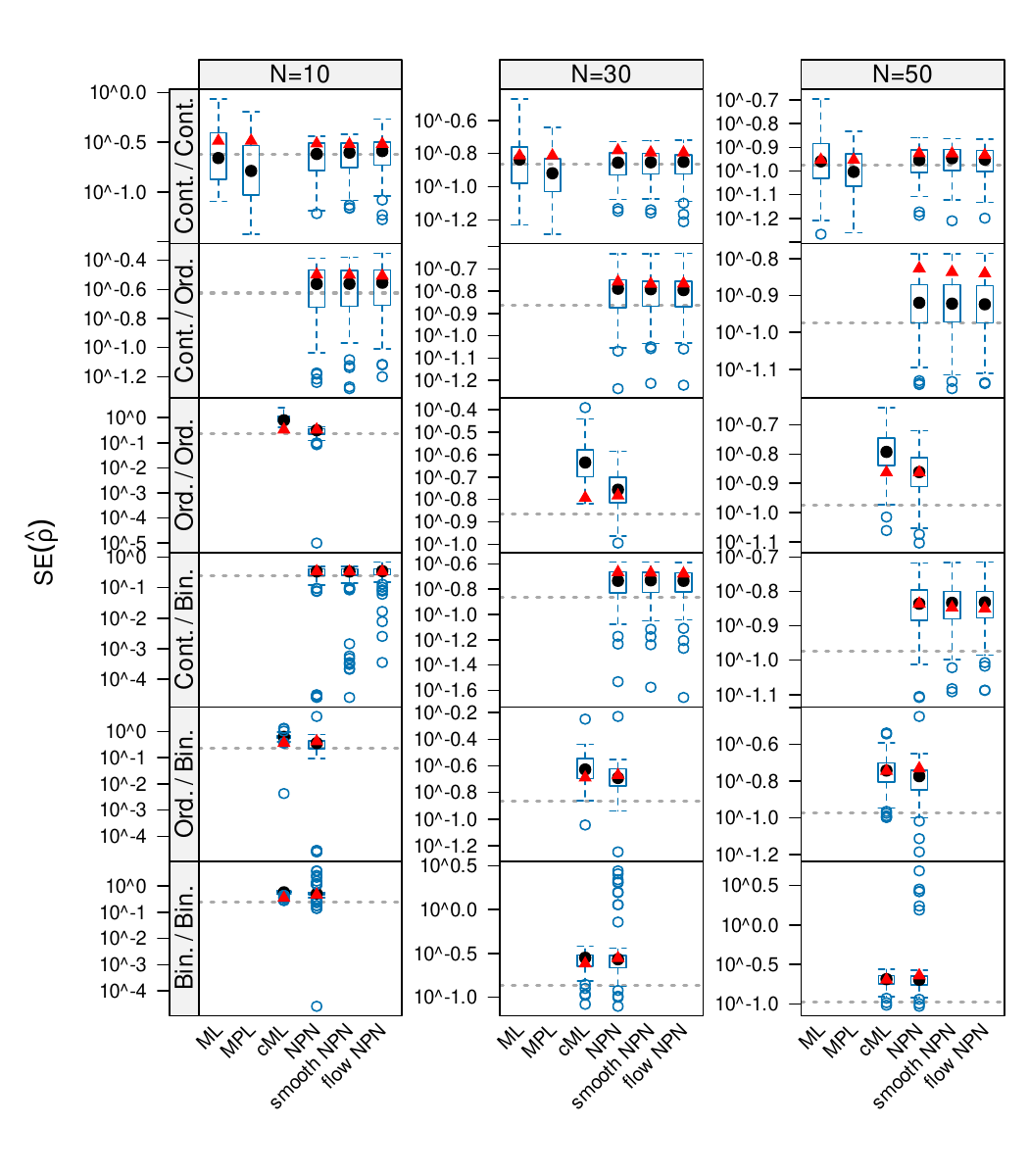} 
\end{knitrout}
\end{center}
\caption{Bivariate polychoric correlations: Distribution of $100$ standard errors for
estimators of $\rho = 0.5$ from $N$ bivariate observations measured at different scales: continuous (Cont.), ordinal
(five levels, Ord.), and binary (Bin.).  For continuous variables, maximum likelihood
(ML) and two-step maximum pseudo likelihood (MPL) approaches are shown as
competitors, for categorical variables, the composite maximum likelihood (cML)
is presented. The horizontal line indicates the semiparametric efficiency
bound and red triangles the standard deviation of $\hat{\rho}_1, \dots,
\hat{\rho}_{100}$. \label{fig:SErho}}
\end{figure}

For a true $\rho = 0.5$, the distribution of the estimators and their
standard errors are given in Figures~\ref{fig:rho} and \ref{fig:SErho}.
For continuous variables, all three flavours of the \NPN attain the
semiparametric efficiency bound for all sample sizes and the corresponding
$\Delta$ standard errors are very close to the theoretical value. In
comparison for small
sample sizes, the classical two-step maximum pseudo likelihood 
copula estimators are slightly more biased, this
also applies to their standard errors. For larger sample sizes, the
estimated correlations are identical, unlike the corresponding standard
errors. 

When at least one variable is
categorical, the NPN estimators are still unbiased but the variance of the
estimators increases slightly. Especially for small sample sizes, the standard errors by the
competing procedures under- or over-estimate the true variation, whereas the
standard errors obtained from inverting the Hessian of some \NPN reflect the
variability of the corresponding estimates closer. For larger sample sizes,
these differences become very small. For smaller and larger values of
$\rho$ in the data generating process, results are given in
\SM~\ref{app:poly}. Especially for very high correlations and when both
variables are binary, the estimation performance as well as the quality of
the standard errors degrades.

\added{In the multivariate setup with $J = 3, 6, 9$ ordinal outcomes, we
compared the polychoric correlations estimated by maximising the composite
likelihood with nonparanormal estimates. The data generating process was the
same as in the bivariate case, except that we randomly sampled $\lambdavec$
from uncorrelated uniform distributions on $[-1,1]$ for each of the $100$ simulation
runs. We report the root mean squared error between the true and estimated
$\nicefrac{J(J-1)}{2}$ polychoric correlations as well as computing times 
in \SM~\ref{app:mpoly}. With increasing dimension and increasing sample
size, the number of convergence failures increases for cML, whereas
optimisation of the \NPN was always possible. In case of successful
convergence, the root mean squared errors were comparable. Availability of
the score function for the \NPN resulted in faster computing times compared to
optimisation of the composite likelihood relying on numerically approximated
scores.}

\section{Application Domains} \label{sec:appl}

The range of applications of the NPN model, especially with
covariate-adjustment, is quite broad and we sketch possible parameterisations
for some interesting applications in this section. We start with probit
models for multivariate binary outcomes \citep{Lesaffre_Kaufmann_1992}, that
is, $\rY_j \in \{0, 1\}$ for all $j = 1, \dots, J$. The marginal
distributions are given by 
$\h_j(\ry_j \mid \rx) = \eparm_{j1} + \rx^\top \shiftparm_j$, where
$\eparm_{j1}$ is the intercept term for the $j$th binary response.
This concept was later generalised to ``Copula regression'' via a multivariate GLM
formulation \citep{Song_Li_Yuan_2009,Masarotto_Varin_2012}. For example,
marginal binary logistic models feature
$\h_j(\ry_j \mid \rx) = \Phi^{-1}(\text{expit}(\eparm_{j1} + \rx^\top \shiftparm_j))$
as marginal transformation functions.
The notion of ``polychoric correlations'' \citep{Joergeskog_1994} was
extended to multivariate proportional-odds models 
\citep[\eg][]{Hirk_Hornik_Vana_2019}. For ordered sample spaces $\samY_j = \{\upsilon_{j1} < \cdots <
\upsilon_{jK(j)}\}$, a marginal proportional-odds model corresponds to the
transformation
$\h_j(\ry_j \mid \rx) = \Phi^{-1}(\text{expit}(\eparm_{jk} + \rx^\top
\shiftparm_j))$. In all these models, the \NPN
$\ell(\parm, \lambdavec)$ can be maximised simultaneously in all model
parameters. This also allows likelihood inference for contrasts of
marginal parameters, for example when the hypothesis $\eshiftparm_{j1} = 0$
for all $j = 1, \dots, J$ is of interest.

For counts $\rY_j \in \NN$, \cite{Siegfried_Hothorn_2020} suggested marginal
proportional-odds models $\h_j(\ry_j \mid \rx) =
\Phi^{-1}(\text{expit}(\basisy_j(\lfloor \ry_j \rfloor)^\top \varparm_j +
\rx^\top \shiftparm_j))$ which, for multiple count outcomes, can be
estimated jointly by maximising the \sNPN $\ell(\parm(\varparm),
\lambdavec)$ \citep{Graz_Barbanti_Brandl_2026}.  This also generalises the bivariate count models by
\cite{Niehaus_Zhu_Cook_2024}.  For continuous outcomes, \cite{Mai_Zou_2014}
and \cite{Sewak_Siegfried_Hothorn_2026} studied transformation discriminant analysis models,
where marginal transformations include shift and possibly scale effects
differentiating between two (or more) classes, see also
Section~\ref{subsec:DA} for a worked example involving both the \fNPN and
\mNPN.  The latter likelihood is relevant for the estimation of graphical
models \citep{Goebler_2024} or structural equation models
\citep{Pritikon_Brick_Neale_2018} for mixed outcomes, for the generation
of synthetic data from such models \citep[\eg for missing value
imputation,][]{Zhao_Udell_2020,Christoffersen_2021}, and for prognostic
covariate adjustment under non-collapsibility \citep{Dandl_Hothorn_2026}.

In survival analysis, multivariate survival times can be analysed by
NPN models with Cox-type margins. Independent censoring
requires the application of the \mNPN. Such models
have been suggested for the case $J = 2$ \citep{Marra_Radice_2018,Ding_Sun_2022}, 
NPN models also allow $J > 2$. A very important topic are recently suggested models 
for dependent censoring. In addition to some time to event of interest $T >0$, one
also observes drop-out times $C > 0$ and times of administrative independent
censoring $A > 0$. For each subject, only $\min(T, C, A)$ can be observed.
\cite{Czado_2022} and \cite{Deresa_Keilegom_2024} proved that the
parameters of a suitably defined NPN model are identified even
under this partial information. More specifically, with
$\h_T(t \mid \rx) = \Phi^{-1}(\text{cloglog}^{-1}(\basisy_T(t)^\top \varparm_T + \rx^\top \shiftparm_T))$
and $\h_C(c \mid \rx) = \Phi^{-1}(\text{cloglog}^{-1}( \basisy_C(c)^\top \varparm_C + \rx^\top \shiftparm_C))$
one can estimate the latent correlation $\rho = \nicefrac{-\lambda_{21}}{\sqrt{1 +
\lambda_{21}^2}}$ if $\basisy_C(c)^\top = (1, \log(c))$, that is, when the marginal drop-out
time follows a Weibull model. The marginal time to event might even follow a 
Cox proportional hazards model, for example with marginal log-baseline cumulative
hazard function $\basisy_T(t)^\top \varparm_T$ parameterised in terms of a
polynomial in Bernstein form. Both marginal distributions ensure that covariate effects are
interpretable as marginal
log-hazard ratios. The log-likelihood for an observed event time
(\ie $T = t, C > t$)
is the \mNPN $\tilde{\ell}_{C \mid T}(\varparm_T, \parm_C(\varparm_C),
\lambda_{21})$. For
a drop-out (\ie $T > c, C = c$), the log-likelihood is the \mNPN
$\tilde{\ell}_{T \mid C}(\varparm_C, \parm_T(\varparm_T), \lambda_{21})$.
Administratively censored subjects (\ie $T > a, C > a$) further add
the \NPN $\tilde{\ell}_2(\parm_T(\varparm_T), \parm_C(\varparm_C),
\lambda_{21})$, \added{an illustration is given by \cite{Siegfried_2026}}.

Finally, all the models above can be coupled with covariate-dependent copula
parameters~(\ref{fm:Lx}) as explained in
\cite{Klein_Hothorn_Barbanti_2020} and \cite{Barrathh_Boyd_2023}, for example when
estimating time-varying graphical models \citep{Lu_Kolar_Liu_2018}. 
Unfortunately, and unlike models with constant $\lambdavec$ parameters, 
such models are in general not invariant to the order in which responses
enter the model.




\section{Discussion}

Given the plethora of inference procedures for many special cases of the
NPN model, one might wonder in which cases
optimisation of the \NPN, or any of the approximations discussed in this
paper, is beneficial.  From a methodological point of view, the \NPN
provides a benchmark against which other approximations, for example the
composite likelihood in multivariate regression models \citep{Nikoloulopoulos_2023}, 
can be evaluated.  We present a simple
version of such a benchmark comparison in Section~\ref{sec:empeval},
comparing the two-step pseudo and composite maximum likelihood approaches to several
flavours of the \NPN, both in terms of their estimation accuracy and
corresponding variability assessment. The exercise shows that \NPN
estimators exhibit the variability of semiparametric efficient
estimators for at least ordered response variables. For small samples, the
estimation accuracy of \NPN estimators is better, compared to two-step
estimators. Practically
even more relevant is the improved accuracy of standard
errors obtained from the \NPN approaches. On these grounds, the application of standard
maximum-likelihood inference procedures to more complex models, for example 
when one is interested in supplementing bootstrap procedures by 
simple Wald tests and confidence intervals in dependent censoring models
\citep{Deresa_Keilegom_2024}, seems promising.

An important contribution is the ability to estimate models when the
response types are mixed, that is, some variables can be considered
continuous while others are clearly discrete. The same applies to missing
values in some of the response variables.
The \NPN allows a straightforward handling of observations missing at
random. We simply use the datum $(-\infty, \infty)$ when computing the contribution of the
$j$th, missing, covariate to the likelihood. Imprecise
measurements can be handled via interval-censoring.

From a more theoretical point of view, the consistency of the two-step pseudo maximum likelihood approach, 
based on normal or winsorised scores, in combination with the graphical lasso was recently demonstrated 
in ultra-high dimensions by \cite{Mai_He_Zou_2023}. So far, such a result is
only available when all responses are absolutely continuous and in the
absence of any additional parameters in the marginal or joint distributions.
The non-convexity of the negative nonparanormal log-likelihoods
studied here renders them unattractive for penalisation approaches in higher
dimensions. However, the contribution might still be
useful for the estimation of graphical models for non-normal and potentially
discrete responses in high-dimensions. Following \cite{Xui_Zou_2012} or
\cite{Suggala_Yang_Ravikumar_2017}, bivariate NPN models could be employed to estimate the polychoric
correlations $\rho_{j\jmath}$. The matrix $(\hat{\rho}_{j\jmath})_{1 \le j < \jmath \le
J}$ with $\hat{\rho}_{jj} \equiv 1$ can then replace the sample covariance matrix in
a graphical lasso, neighbourhood Dantzig selector, or CLIME. \cite{Xui_Zou_2012}
demonstrated that their ``rank-based'' versions are consistent with the same
rates of convergence as the original versions based on the sample covariance
matrix of normal data.

Efficiency results on such two-step estimators are sparse.
\cite{Klaassen_Wellner_1997} demonstrated efficiency of the correlation
parameter in a bivariate Gaussian copula, and we utilised this ground truth in the
simulation experiments in Section~\ref{sec:empeval}. Even in this simple
case, the two-step marginal distributions are inefficient. 
The \fNPN is conceptually very similar to the semiparametric efficient
sieve maximum-likelihood estimators studied by
\cite{Chen_Fan_Tsyrennikov_2006}.  The main difference lies in their choice
of a sieve approximation for marginal densities whereas we utilise
polynomials in Bernstein form $\basisy_j(\ry_j)^\top \varparm_j$ to
approximate marginal transformation functions.  If one allows the number of
basis functions $K(j)$ in $\basisy_j$, and therefore the number of
coefficients $\varparm_j$, to depend on the same size $N$, the sieve space
proposed and analysed by \cite{McLain_Ghosh_2013} emerges. \cite{Chen_Fan_Tsyrennikov_2006}
also proved that semiparametric efficiency carries over to models where some
of the marginal distributions are fully parametric. These results make the
NPN model, its smooth parameterisation and the corresponding
\fNPN, especially when coupled with the ACS optimisation method,
promising candidates for future 
research.

\section{Data, Script and Code Availability} \label{sec:comp}

All computations were performed using \textsf{R} version 4.6.1 
\citep{R}. All flavours of nonparanormal log-likelihoods were computed using infrastructure in
package \pkg{mlt} \citep{pkg:mlt} based on algorithms for the evaluation
of multivariate normals in \pkg{mvtnorm} \citep{pkg:mvtnorm}.
The convex parameterisation of the LDA model was estimated by package
\pkg{CVXR} \citep{pkg:CVXR}. Maximum pseudo and composite likelihood estimates of
polychoric correlations in Section~\ref{subsec:poly} were computed using
packages \pkg{copula} \citep{pkg:copula}, with variant ``ML'' corresponding
to \code{fitCopula(..., type = "ml")} and variant ``MPL'' to 
\code{fitCopula(..., type = "mpl")},  and \pkg{mvord} \citep{pkg:mvord},
respectively. A low-level interface to different forms of the nonparanormal
log-likelihoods is available from package \pkg{mlt} \citep{pkg:mlt} via the
\code{mlt::mmlt} function. Package \pkg{tram} provides a high-level
interface as \code{tram::Mmlt}. 

The location-scale transformation discriminant analysis model for HCC
diagnosis under limits-of-detection was estimated by the following code
\begin{Schunk}
\begin{Sinput}
> library("tram")
> ### run demo("npn") from tram package for full reproducibility
> 
> ### marginal location-scale models
> mDKK <- BoxCox(
+     DKK ~                               ### probit, h(DKK) via Bernstein
+     x                                   ### location non-HCC / HCC
+     | x,                                ### scale non-HCC / HCC
+     data = HCC)
> mOPN <- BoxCox(OPN ~ x | x, data = HCC)
> mPIV <- BoxCox(R(	
+     Surv(PIV, event = PIV < PIVm),      ### right censoring
+     as.R.interval = TRUE) ~             ### empirical likelihood
+     x | x,                              ### location-scale
+     data = HCC)
> mAFP <- BoxCox(R(Surv(AFP, event = AFP < AFPm), as.R.interval = TRUE) ~ 
+                x | x, data = HCC)
> ### joint estimation of marginal and Gaussian copula parameters, s = 2
> ### location-scale transformation discriminant analysis
> m <- Mmlt(mDKK, mOPN, mPIV, mAFP, data = HCC)
> ### marginal parameters
> coef(m, type = "marginal")
> ### copula parameter: Lambda
> coef(m, type = "Lambda")
> ### standard errors for all parameters
> sqrt(diag(vcov(m)))
> ### convex approximations
> ## pseudo
> mm <- Mmlt(mDKK, mOPN, mPIV, mAFP, data = HCC, domargins = FALSE)
> ## sequential
> ms <- Mmlt(mDKK, mOPN, mPIV, mAFP, data = HCC, sequentialfit = TRUE)
\end{Sinput}
\end{Schunk}

The complete \proglang{R} code necessary to reproduce empirical results discussed in Section~\ref{sec:empeval} 
is provided in
\begin{itemize}
\item \pkg{mvtnorm}: \href{https://doi.org/10.32614/CRAN.package.mvtnorm}{10.32614/CRAN.package.mvtnorm}
\item \pkg{mlt}: \href{https://doi.org/10.32614/CRAN.package.mlt}{10.32614/CRAN.package.mlt}
\item \pkg{tram}: \href{https://doi.org/10.32614/CRAN.package.tram}{10.32614/CRAN.package.tram},
including data and code for Section~\ref{subsec:DA} in \\
\code{tram/inst/npn\_empeval} and for Section~\ref{subsec:poly} in
\code{tram/inst/npn\_simulation}.
\end{itemize}

\section{Supplementary Information} \label{app:impl}

\begin{itemize}
\item Appendix~\ref{app:proofs}, Proofs: Proof of Theorem~\ref{thm:fNPN} and Theorem~\ref{thm:NPN}.
\item Appendix~\ref{app:poly}, Polychoric Correlations: Additional
simulation results for the bi- and multivariate case.
\item Supplementary
Material, \href{https://doi.org/10.32614/CRAN.package.mvtnorm}{10.32614/CRAN.package.mvtnorm}:
\href{https://CRAN.R-project.org/web/packages/mvtnorm/vignettes/lmvnorm_src.pdf}{Likelihood and Score
Functions}:
A modular re-implementation of \cite{Genz_1992} algorithm tailored to the
evaluation of the different nonparanormal log-likelihoods. Implementation aspects of mixed continuous and
discrete normal log-likelihoods are discussed in Chapter 5.
Log-likelihoods for the case $s = 2$ are described in
Chapter 6. The chain-rule to derive scores with respect to $\mLambda^{-1}$ 
is given in Section~3.2.
\end{itemize}

\section*{Funding}

Financial support by Swiss National Science Foundation \textnumero~200021\_219384 is acknowledged.

\section*{Conflict of Interest Disclosure}

The author declares he has no conflict of interest relating to the content of
this article.

\newpage

\bibliography{mlt,packages}

\newpage

\begin{appendix}

\section{Proofs} \label{app:proofs}

Theorem~\ref{thm:fNPN}.
\begin{proof}
As a function of $\zvec \in \RR^J$, $\lVert \mLz^{(s)}(\lambdavec) \zvec
\rVert_2^2$ is convex for $s = 1, 2$ and each fixed $\lambdavec \in \RR^{\nicefrac{J(J - 1)}{2}}$. 
The argument is completed noting that $\rz_j = \basisy_j(\ry_j)^\top \varparm_j$
is linear in $\varparm_j$.

As a function of $\lambdavec \in \RR^{\nicefrac{J(J - 1)}{2}}$, 
$\lVert \mLz^{(1)}(\lambdavec) \zvec \rVert_2^2 = \lVert \mLambda \zvec \rVert_2^2$ is convex 
for each fixed $\zvec \in \RR^J$. For $s = 2$, we follow \cite{Khare_2019}
and write
\begin{eqnarray*}
-\tilde{\ell}_{J,i}^{(0)}(\mLz^{(2)}) = -\left(
- \frac{1}{2} \lVert \mLz^{(2)}(\lambdavec) \zvec \rVert_2^2 + \sum_{j = 1}^J
\log\left(\mLz^{(2)}_{jj}\right)\right) = \sum_{j = 1}^J
\frac{1}{2} \lVert \mLz^{(2)}(\lambdavec)_{j\cdot} \zvec \rVert_2^2 -\log\left(\mLz^{(2)}_{jj}\right)
\end{eqnarray*}
as a sum of $J$ independent terms.

For $j = 1$, $\mLz^{(2)}(\lambdavec)_{11} = 1$. For $j = 2$, we add the
constraint $\mLz^{(2)}(\lambdavec)_{22} = \lVert \lambda_{2,\cdot} \lVert_2 = \sqrt{1 +
\lambda_{21}^2}$. This constraint is convex in $\lambda_{21}$ and can be
relaxed to the convex inequality constraint $\lVert \lambda_{2,\cdot} \lVert_2 -
\mLz^{(2)}(\lambdavec)_{22} \le 0$.
For $j > 3$, we write $\mLz^{(2)}(\lambdavec)_{j,\jmath} = \lambda_{j,\jmath}
\mLz^{(2)}(\lambdavec)_{\jmath,\jmath}$ and note that
$\mLz^{(2)}(\lambdavec)_{\jmath,\jmath}$ only depends on $\lambda_{11}, \dots,
\lambda_{\jmath,\jmath - 1}$ for $\jmath = 1, \dots, j - 1$.
Thus, the constraint
$\lVert\lambda_{j,\cdot}\rVert_2 - \mLz^{(2)}(\lambdavec)_{jj} \le 0$ is a
convex inequality constraint. 
\end{proof}
\noindent
Theorem~\ref{thm:NPN}.
\begin{proof}
The normal density $\phivec(\zvec \mid \mLz)$ is log-concave in $\zvec \in \RR^J$
for fixed $\mLz$ and probabilities thereof are again log-concave \citep{Prekopa_1973}.

In a counter-example with $J = 2$ and $s = 1$, we have
\begin{eqnarray*}
\exp\left(\ell^{(s)}_{J,i}(\parm, \lambda)\right) & = & \int\limits_{\eparm_{1,\rk(i,1)-1}}^{\eparm_{1,\rk(i,1)}}
                               \int\limits_{\eparm_{2,\rk(i,2)-1}}^{\eparm_{2,\rk(i,2)}}
    \underbrace{\exp\left(-\frac{1}{2} \rz_1^2\right) \exp\left(-\frac{1}{2}(\rz_2 + \lambda
\rz_1)^2\right)}_{\psi(\lambda)} \, d\rz_2 d\rz_1 \\
\frac{\partial^2 \psi(\lambda)}{\partial^2 \lambda}
& = & 
    \rz_1^2((\rz_2 + \lambda \rz_1)^2 - 1)
      \exp\left(-\frac{1}{2} \rz_1^2\right) \exp\left(-\frac{1}{2}(\rz_2 + \lambda \rz_1)^2\right) 
\end{eqnarray*}
The integral of the latter expression is larger than zero for example for $\lambda = 0$ and
$\eparm_{2,\rk(i,2)-1} > 1$ and thus not concave in $\lambda$ for all
configurations of $\parm$. Similar issues have been noted by
\cite{Lesaffre_Kaufmann_1992} in the multivariate probit model.
\end{proof}

Corollary~\ref{coro:mNPN} follows from Theorem~\ref{thm:fNPN}, noting that $\mC = 1$ and
\begin{eqnarray*}
& & \Phi\left(\eparm_{J,\rk(i, J)} + (\lambda_{J,1}, \dots,
\lambda_{J,J-1})^\top (\h_1(\ry_1 \mid \varparm_1), \dots, \h_\jmath(\ry_{J
- 1} \mid \varparm_{J - 1})\right) - \\
& & \quad \Phi\left(\eparm_{J,\rk(i, J) - 1} + (\lambda_{J,1}, \dots,
\lambda_{J,J-1})^\top (\h_1(\ry_1 \mid \varparm_1), \dots, \h_\jmath(\ry_{J
- 1} \mid \varparm_{J - 1})\right)
\end{eqnarray*}
is log-concave (a probability of a log-concave density and linearity in $\lambda_{J,1}, \dots,
\lambda_{J,J-1})^\top$).

\newpage

\section{Bivariate Polychoric Correlations} \label{app:poly}

Additional simulation results for correlations $\rho \neq 0.5$.

\begin{figure}
\begin{center}
\begin{knitrout}
\definecolor{shadecolor}{rgb}{0.969, 0.969, 0.969}\color{fgcolor}
\includegraphics[width=\maxwidth]{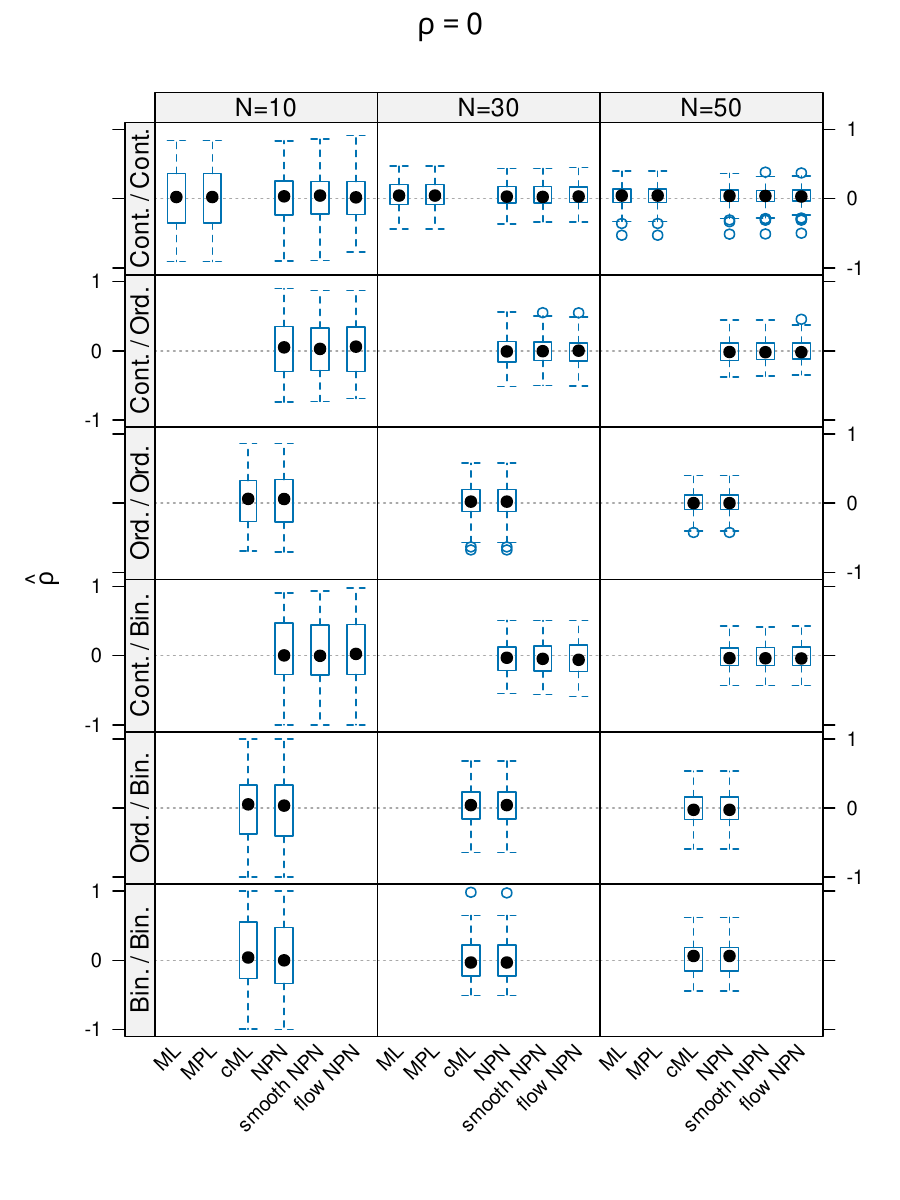} 
\end{knitrout}
\end{center}
\caption{Bivariate polychoric correlations: Distribution of $100$ estimators of $\rho=0$.}
\end{figure}

\begin{figure}
\begin{center}
\begin{knitrout}
\definecolor{shadecolor}{rgb}{0.969, 0.969, 0.969}\color{fgcolor}
\includegraphics[width=\maxwidth]{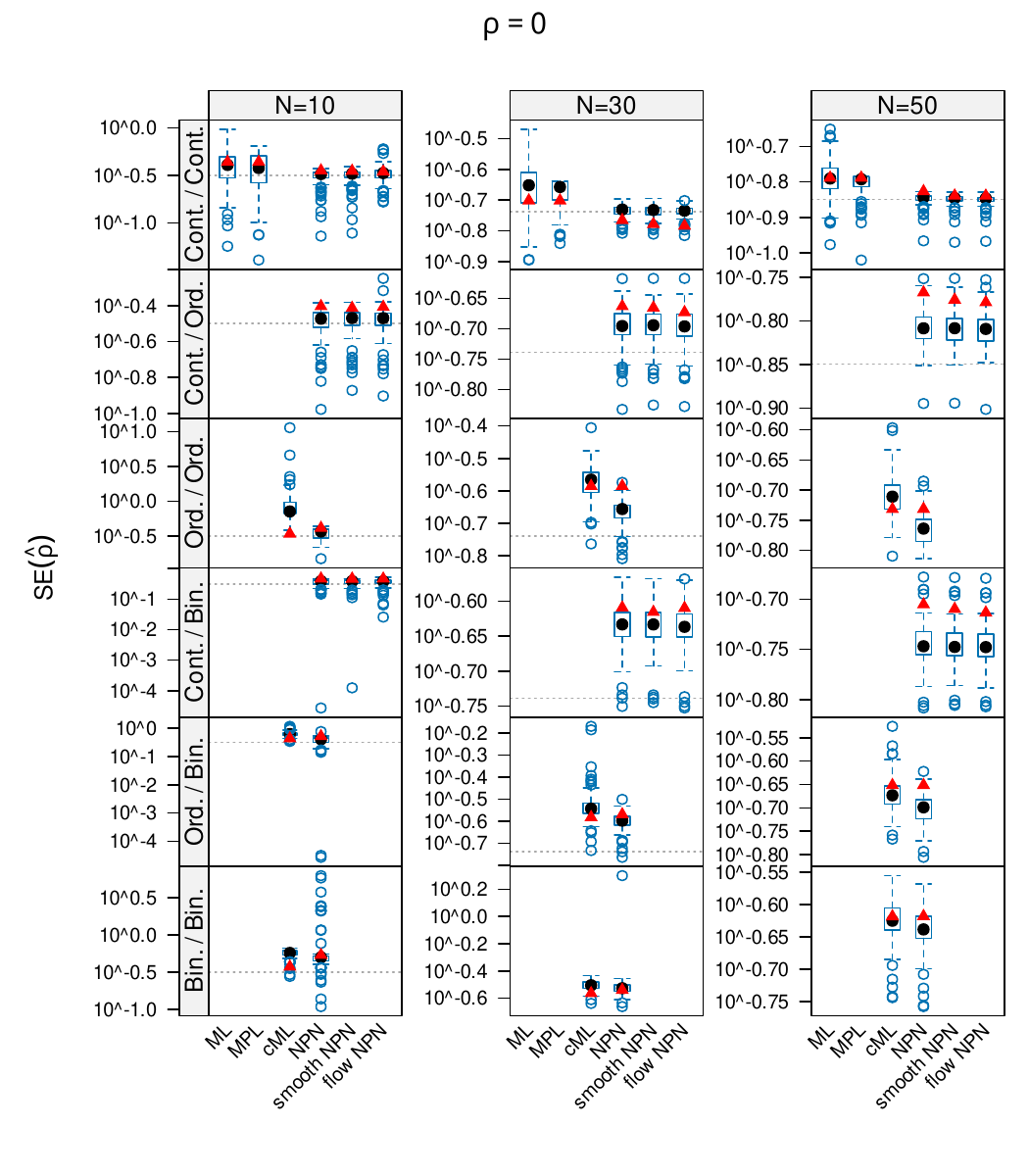} 
\end{knitrout}
\end{center}
\caption{Bivariate polychoric correlations: Distribution of $100$ standard errors for
estimators of $\rho=0$.}
\end{figure}

\begin{figure}
\begin{center}
\begin{knitrout}
\definecolor{shadecolor}{rgb}{0.969, 0.969, 0.969}\color{fgcolor}
\includegraphics[width=\maxwidth]{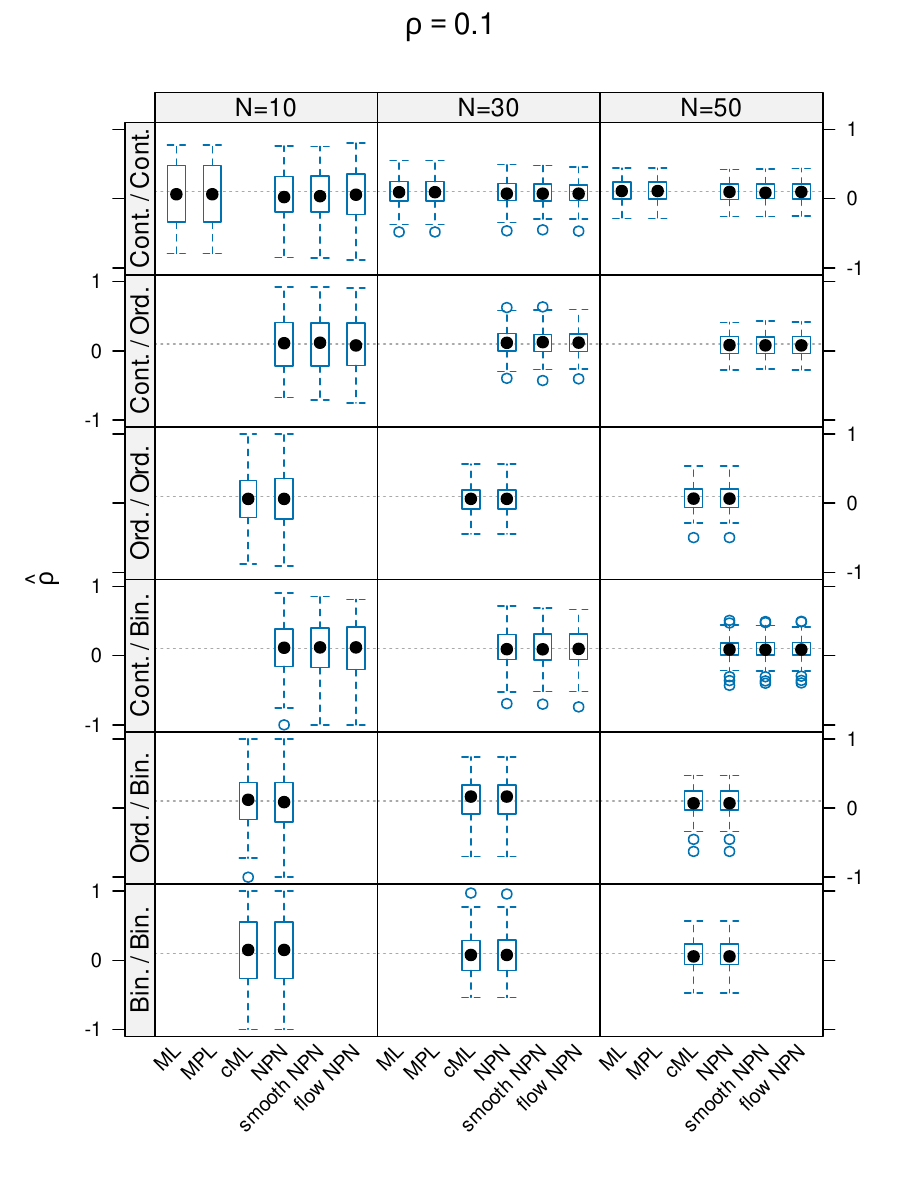} 
\end{knitrout}
\end{center}
\caption{Bivariate polychoric correlations: Distribution of $100$ estimators of $\rho=0.1$.}
\end{figure}

\begin{figure}
\begin{center}
\begin{knitrout}
\definecolor{shadecolor}{rgb}{0.969, 0.969, 0.969}\color{fgcolor}
\includegraphics[width=\maxwidth]{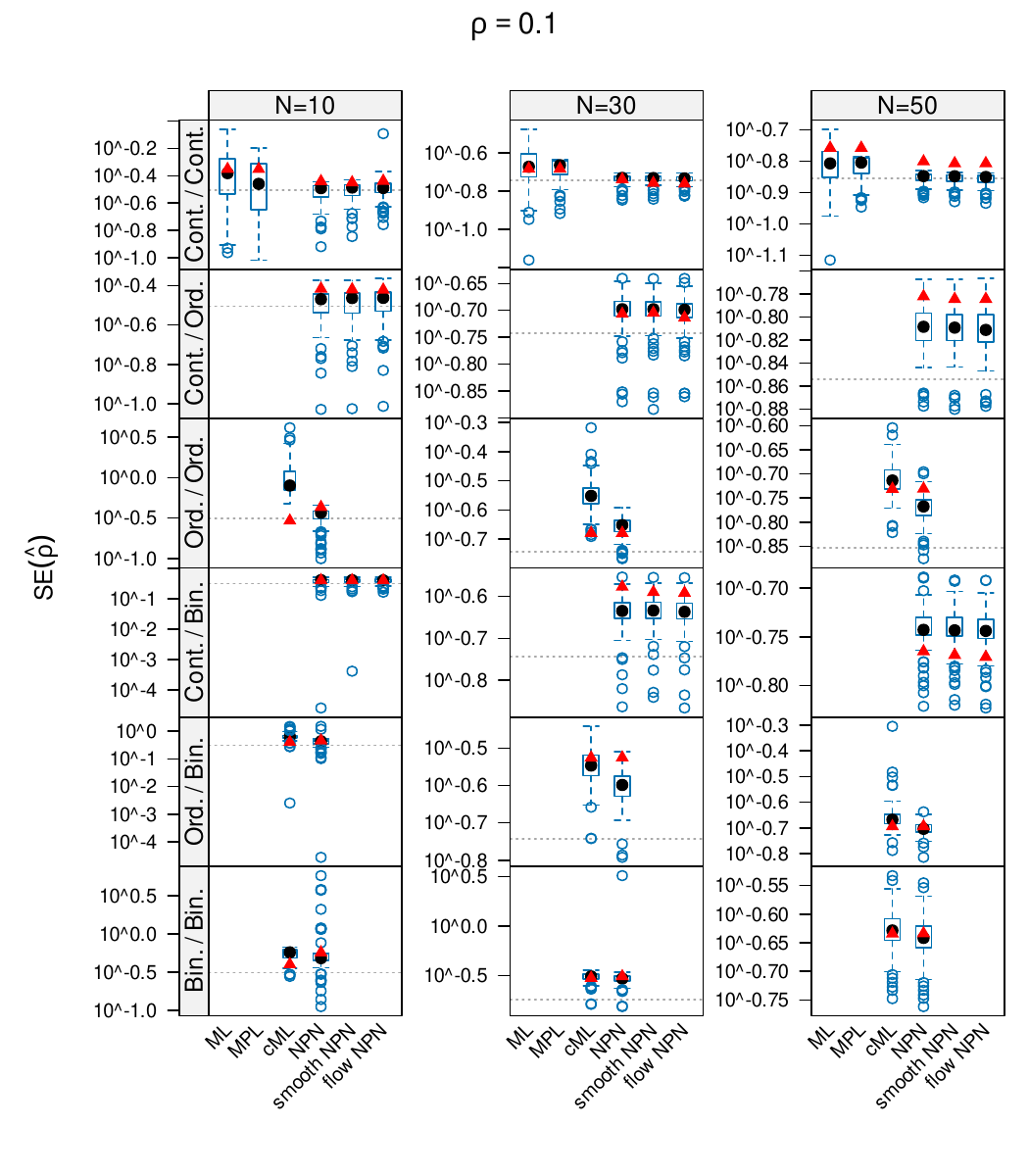} 
\end{knitrout}
\end{center}
\caption{Bivariate polychoric correlations: Distribution of $100$ standard errors for
estimators of $\rho=0.1$.}
\end{figure}

\begin{figure}
\begin{center}
\begin{knitrout}
\definecolor{shadecolor}{rgb}{0.969, 0.969, 0.969}\color{fgcolor}
\includegraphics[width=\maxwidth]{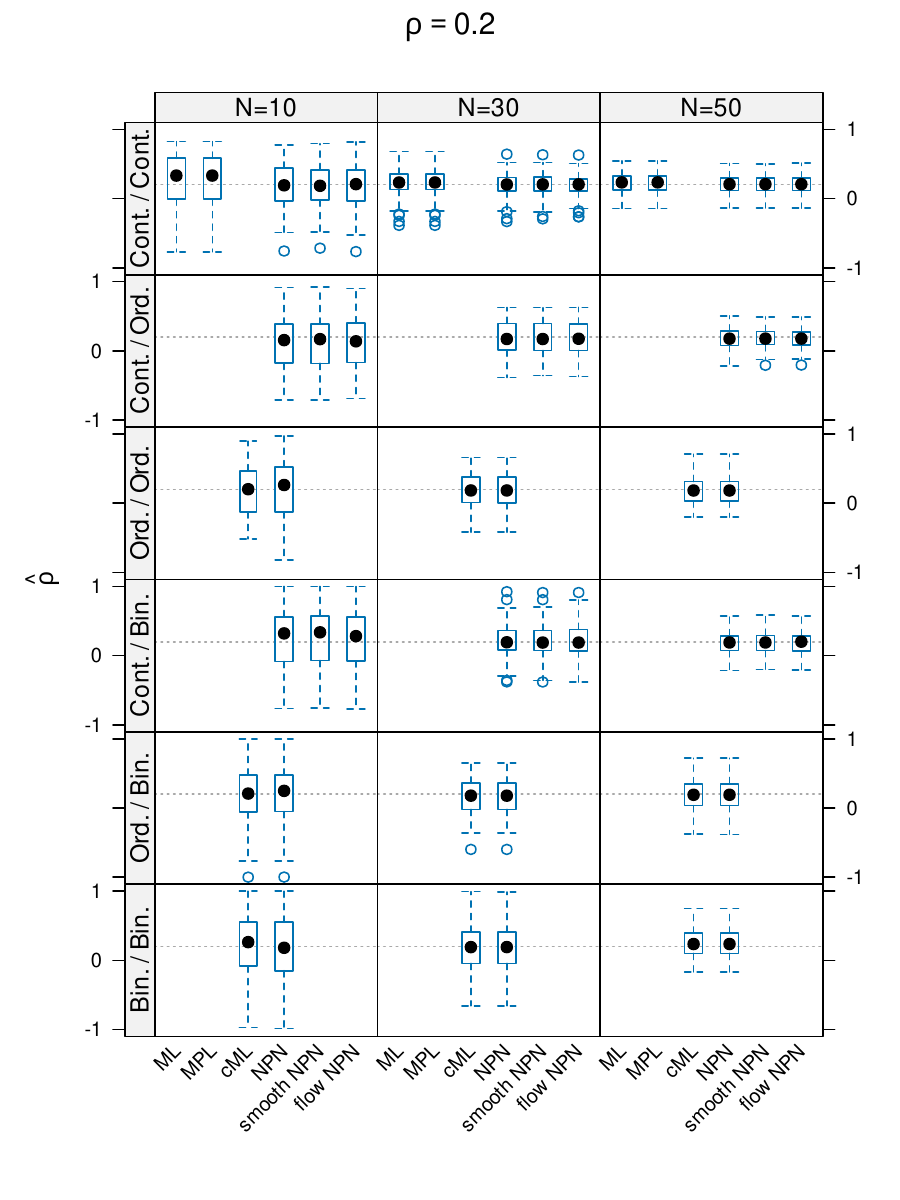} 
\end{knitrout}
\end{center}
\caption{Bivariate polychoric correlations: Distribution of $100$ estimators of $\rho=0.2$.}
\end{figure}

\begin{figure}
\begin{center}
\begin{knitrout}
\definecolor{shadecolor}{rgb}{0.969, 0.969, 0.969}\color{fgcolor}
\includegraphics[width=\maxwidth]{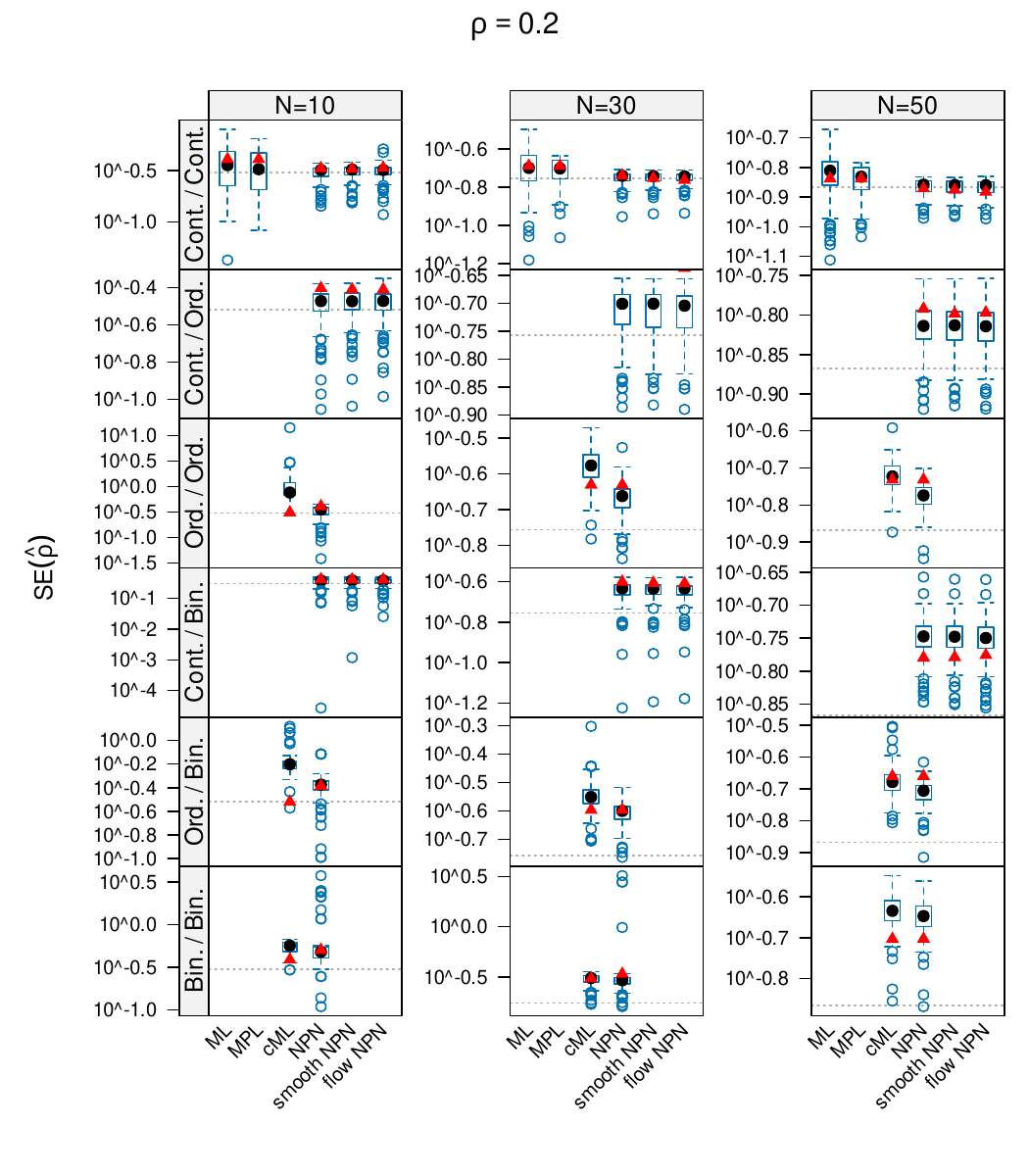} 
\end{knitrout}
\end{center}
\caption{Bivariate polychoric correlations: Distribution of $100$ standard errors for
estimators of $\rho=0.2$.}
\end{figure}

\begin{figure}
\begin{center}
\begin{knitrout}
\definecolor{shadecolor}{rgb}{0.969, 0.969, 0.969}\color{fgcolor}
\includegraphics[width=\maxwidth]{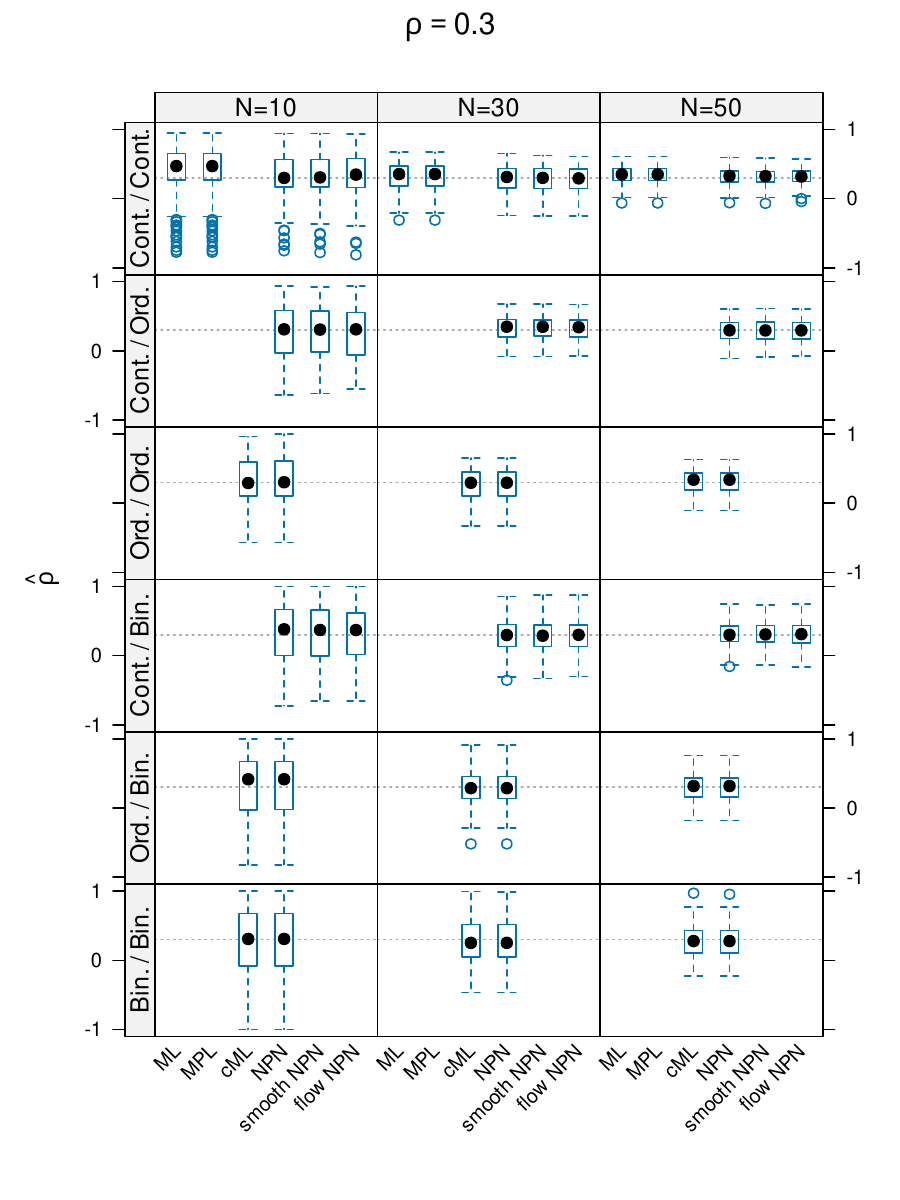} 
\end{knitrout}
\end{center}
\caption{Bivariate polychoric correlations: Distribution of $100$ estimators of $\rho=0.3$.}
\end{figure}

\begin{figure}
\begin{center}
\begin{knitrout}
\definecolor{shadecolor}{rgb}{0.969, 0.969, 0.969}\color{fgcolor}
\includegraphics[width=\maxwidth]{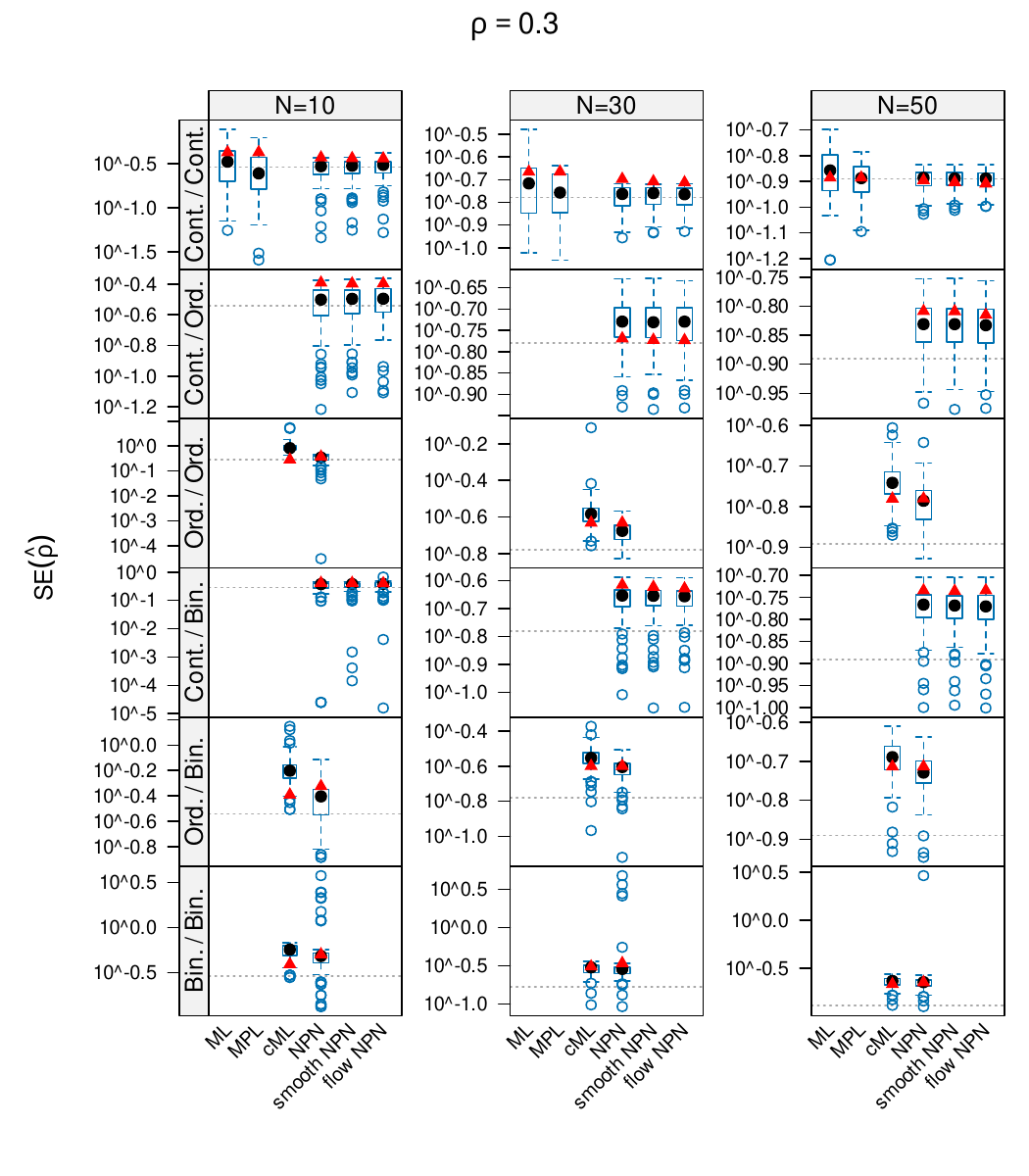} 
\end{knitrout}
\end{center}
\caption{Bivariate polychoric correlations: Distribution of $100$ standard errors for
estimators of $\rho=0.3$.}
\end{figure}

\begin{figure}
\begin{center}
\begin{knitrout}
\definecolor{shadecolor}{rgb}{0.969, 0.969, 0.969}\color{fgcolor}
\includegraphics[width=\maxwidth]{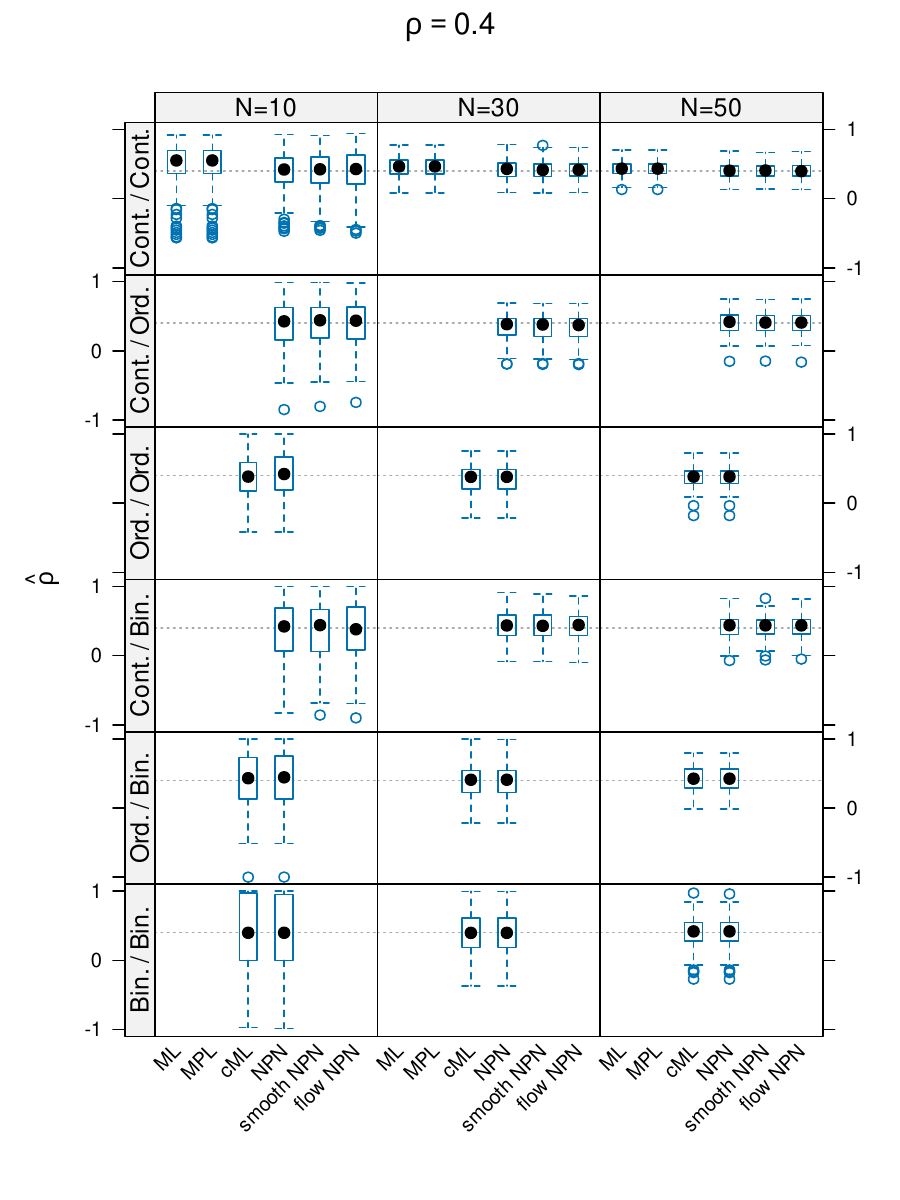} 
\end{knitrout}
\end{center}
\caption{Bivariate polychoric correlations: Distribution of $100$ estimators of $\rho=0.4$.}
\end{figure}

\begin{figure}
\begin{center}
\begin{knitrout}
\definecolor{shadecolor}{rgb}{0.969, 0.969, 0.969}\color{fgcolor}
\includegraphics[width=\maxwidth]{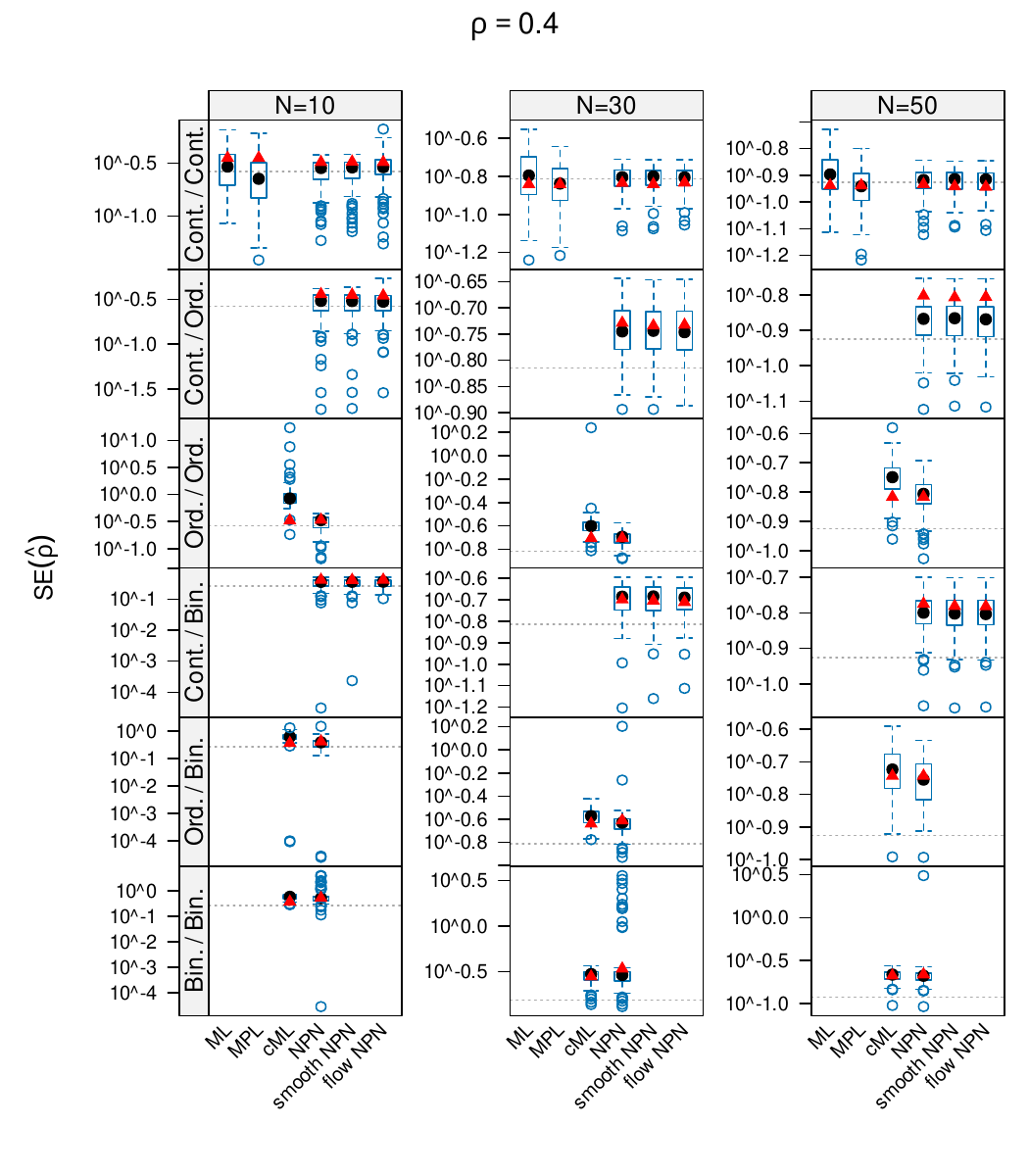} 
\end{knitrout}
\end{center}
\caption{Bivariate polychoric correlations: Distribution of $100$ standard errors for
estimators of $\rho=0.4$.}
\end{figure}

\begin{figure}
\begin{center}
\begin{knitrout}
\definecolor{shadecolor}{rgb}{0.969, 0.969, 0.969}\color{fgcolor}
\includegraphics[width=\maxwidth]{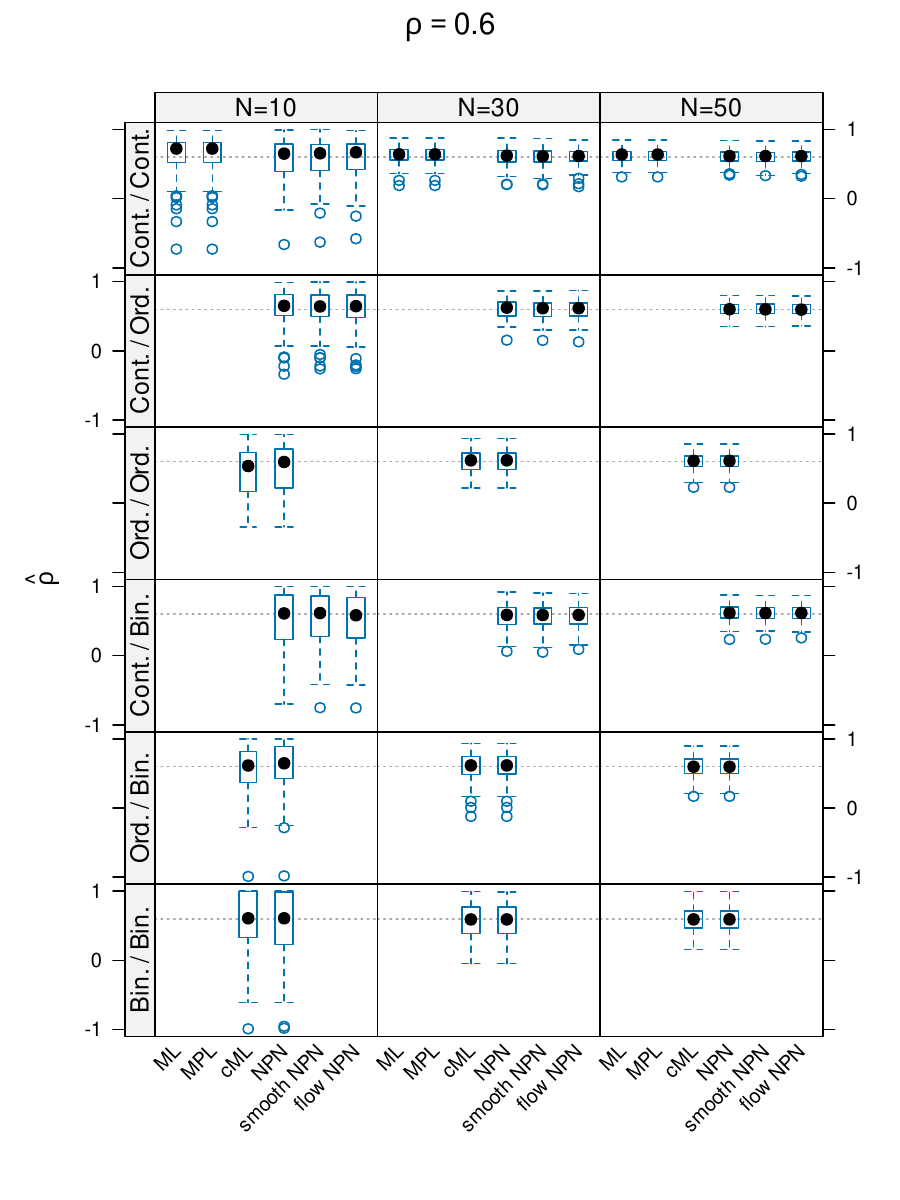} 
\end{knitrout}
\end{center}
\caption{Bivariate polychoric correlations: Distribution of $100$ estimators of $\rho=0.6$.}
\end{figure}

\begin{figure}
\begin{center}
\begin{knitrout}
\definecolor{shadecolor}{rgb}{0.969, 0.969, 0.969}\color{fgcolor}
\includegraphics[width=\maxwidth]{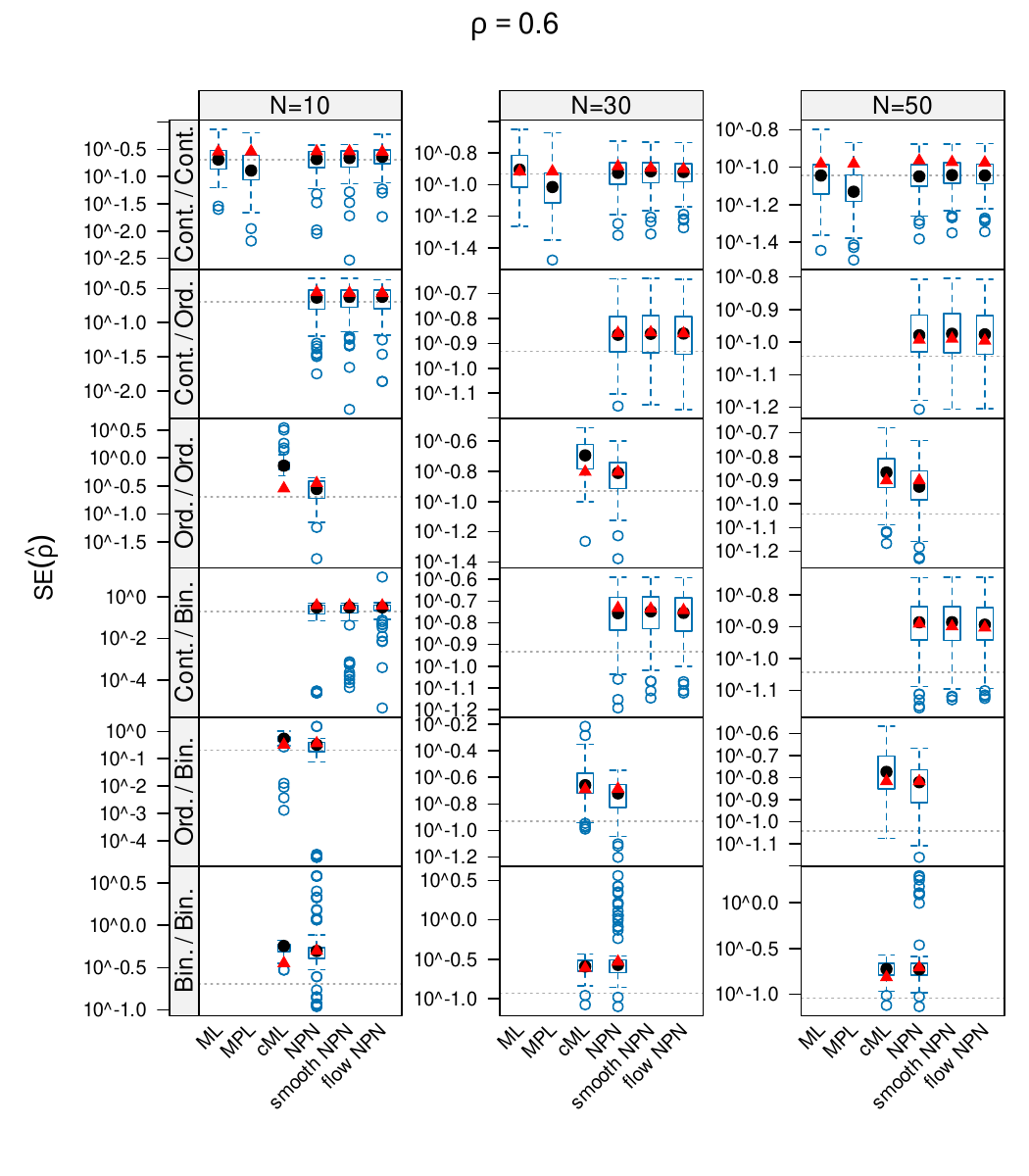} 
\end{knitrout}
\end{center}
\caption{Bivariate polychoric correlations: Distribution of $100$ standard errors for
estimators of $\rho=0.6$.}
\end{figure}

\begin{figure}
\begin{center}
\begin{knitrout}
\definecolor{shadecolor}{rgb}{0.969, 0.969, 0.969}\color{fgcolor}
\includegraphics[width=\maxwidth]{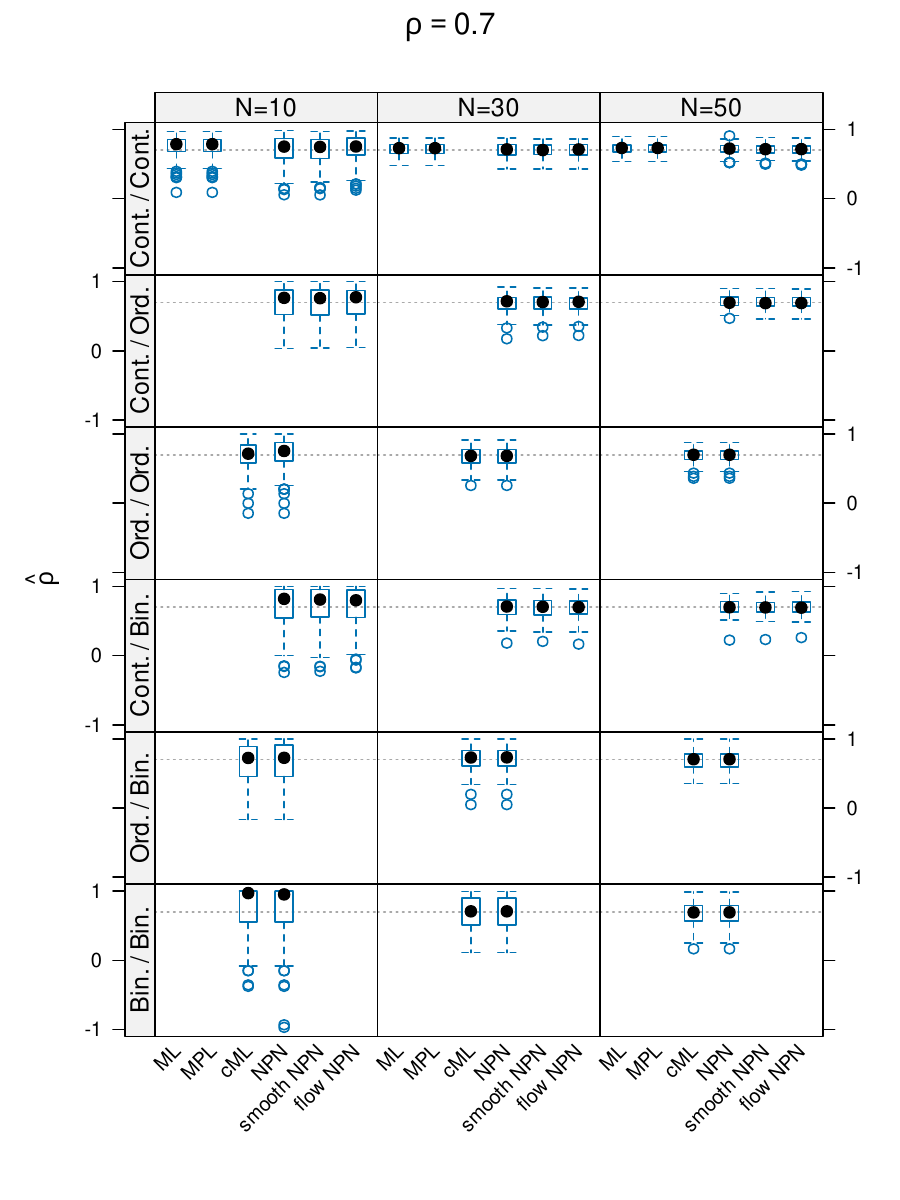} 
\end{knitrout}
\end{center}
\caption{Bivariate polychoric correlations: Distribution of $100$ estimators of $\rho=0.7$.}
\end{figure}

\begin{figure}
\begin{center}
\begin{knitrout}
\definecolor{shadecolor}{rgb}{0.969, 0.969, 0.969}\color{fgcolor}
\includegraphics[width=\maxwidth]{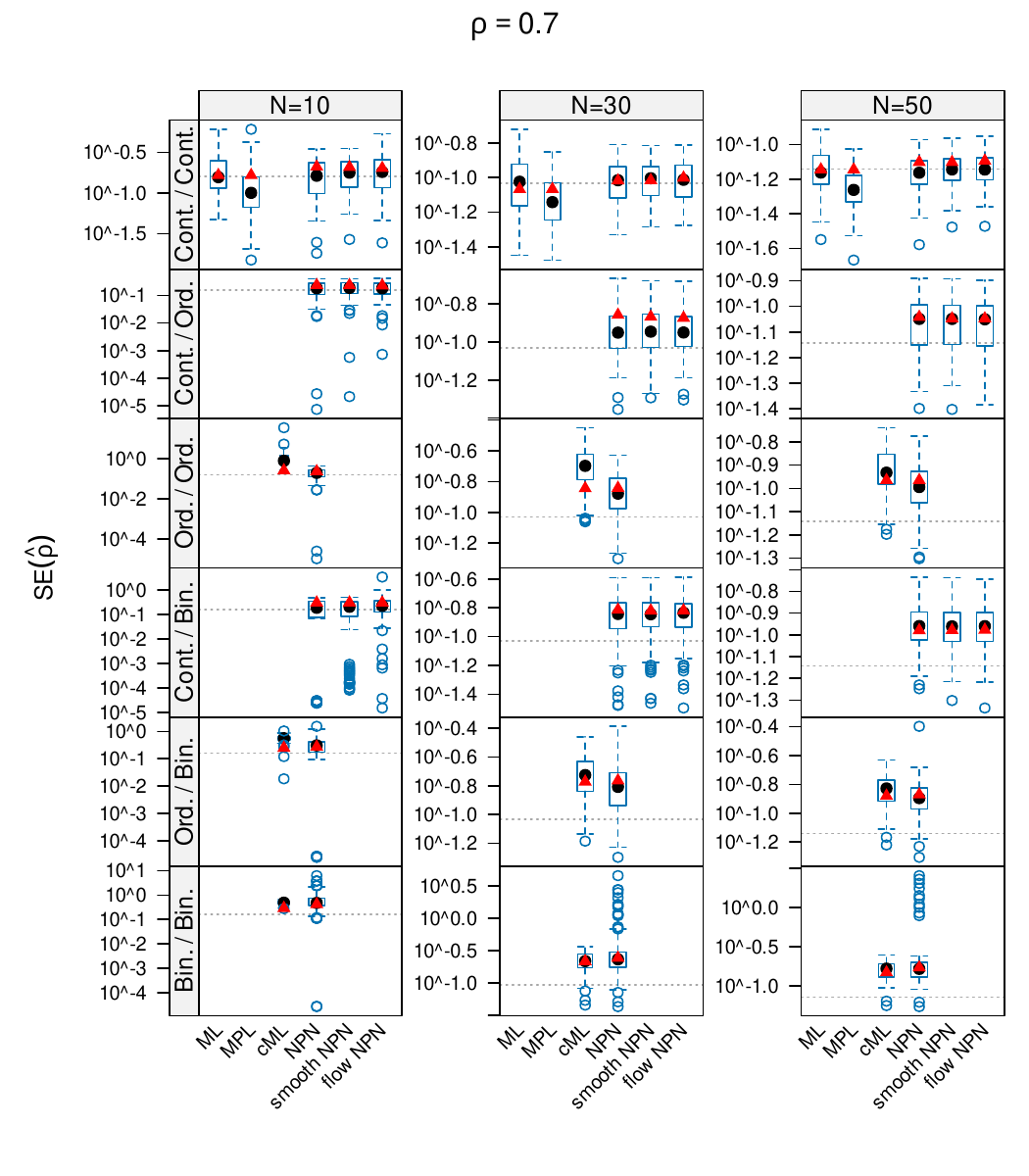} 
\end{knitrout}
\end{center}
\caption{Bivariate polychoric correlations: Distribution of $100$ standard errors for
estimators of $\rho=0.7$.}
\end{figure}

\begin{figure}
\begin{center}
\begin{knitrout}
\definecolor{shadecolor}{rgb}{0.969, 0.969, 0.969}\color{fgcolor}
\includegraphics[width=\maxwidth]{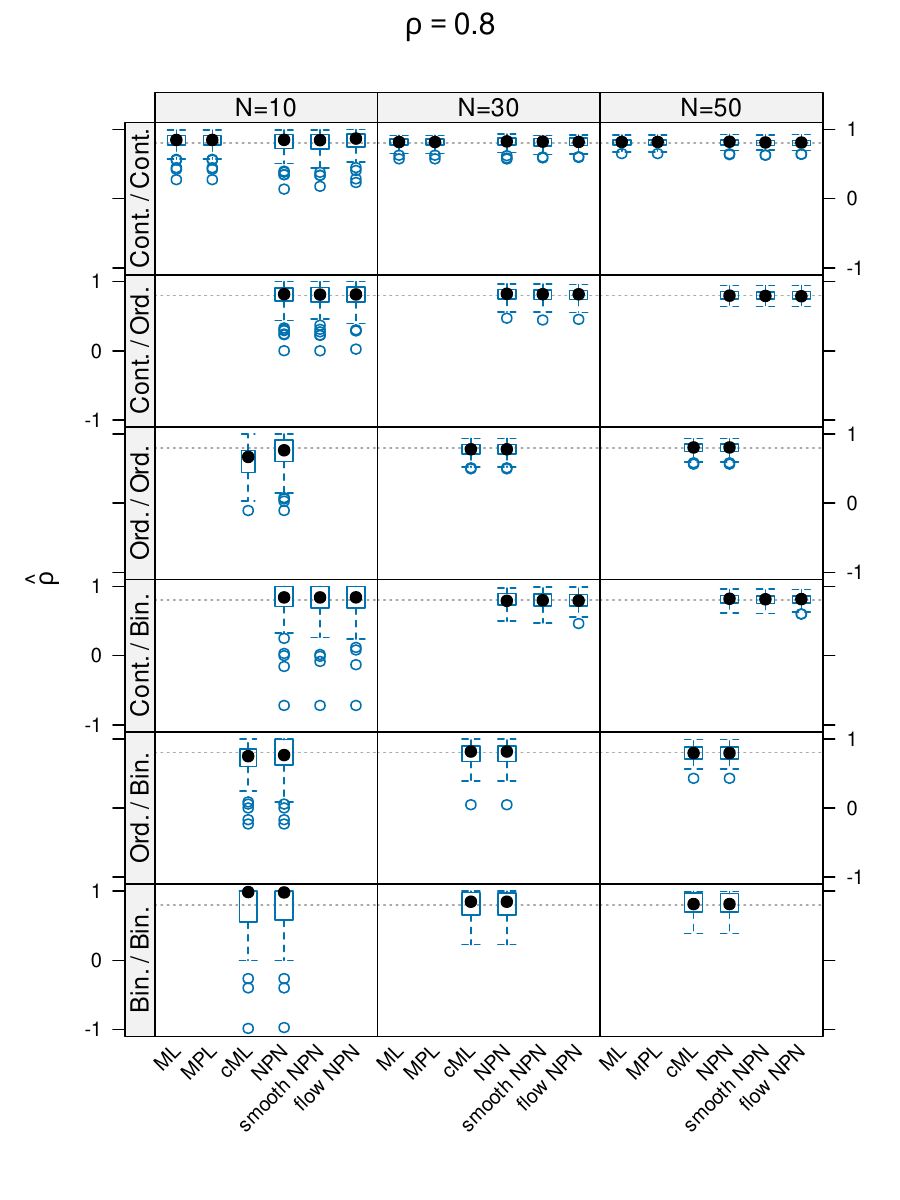} 
\end{knitrout}
\end{center}
\caption{Bivariate polychoric correlations: Distribution of $100$ estimators of $\rho=0.8$.}
\end{figure}

\begin{figure}
\begin{center}
\begin{knitrout}
\definecolor{shadecolor}{rgb}{0.969, 0.969, 0.969}\color{fgcolor}
\includegraphics[width=\maxwidth]{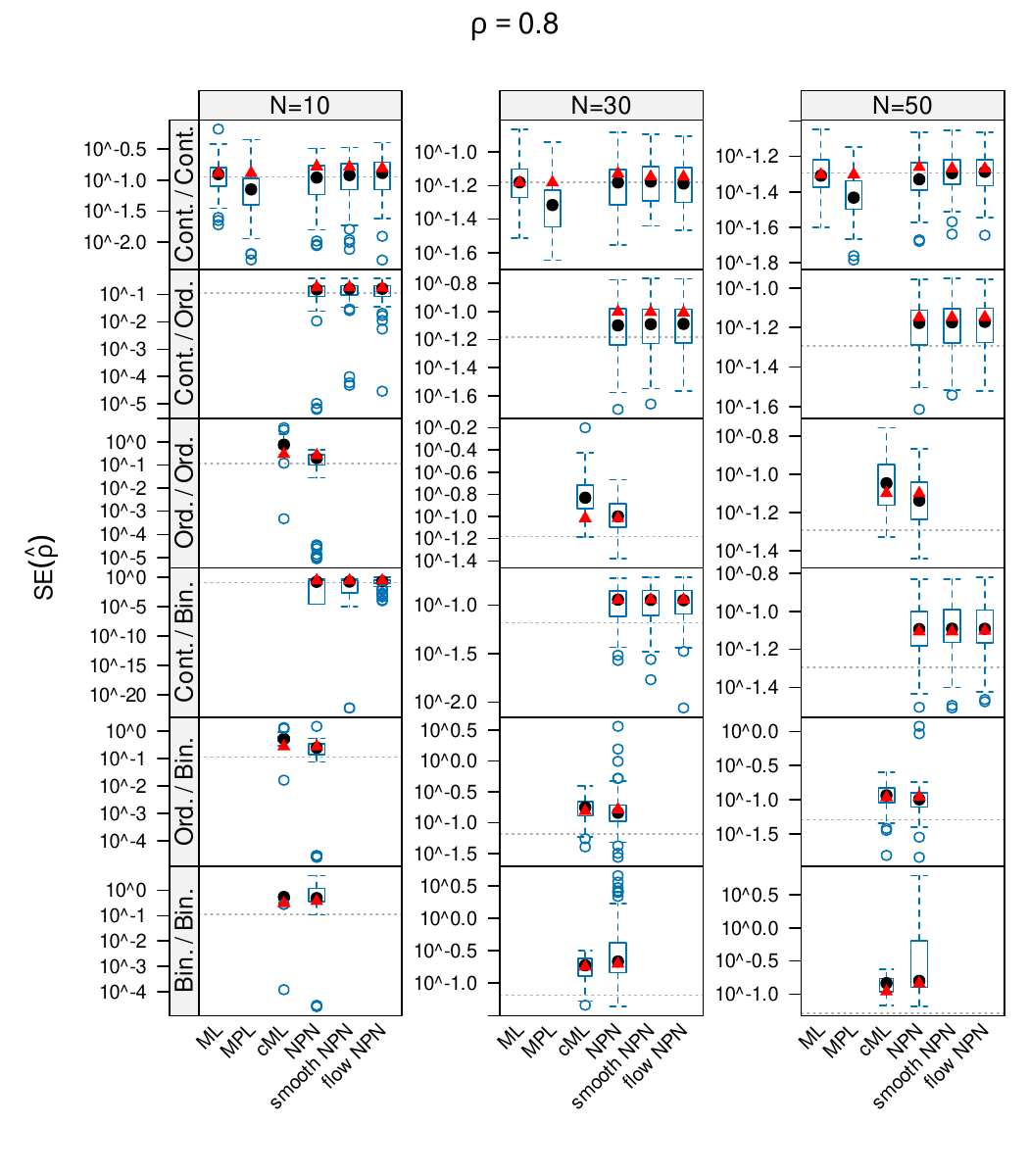} 
\end{knitrout}
\end{center}
\caption{Bivariate polychoric correlations: Distribution of $100$ standard errors for
estimators of $\rho=0.8$.}
\end{figure}

\begin{figure}
\begin{center}
\begin{knitrout}
\definecolor{shadecolor}{rgb}{0.969, 0.969, 0.969}\color{fgcolor}
\includegraphics[width=\maxwidth]{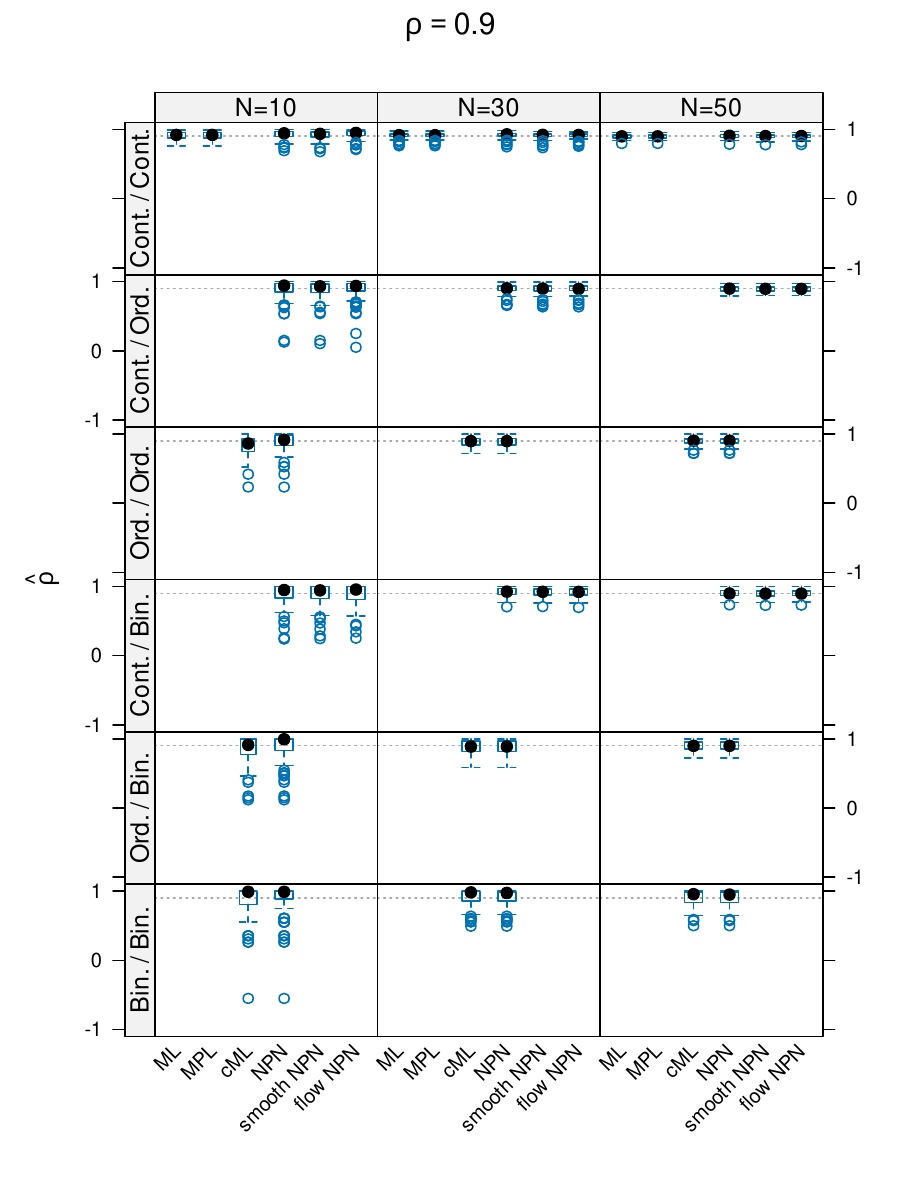} 
\end{knitrout}
\end{center}
\caption{Bivariate polychoric correlations: Distribution of $100$ estimators of $\rho=0.9$.}
\end{figure}

\begin{figure}
\begin{center}
\begin{knitrout}
\definecolor{shadecolor}{rgb}{0.969, 0.969, 0.969}\color{fgcolor}
\includegraphics[width=\maxwidth]{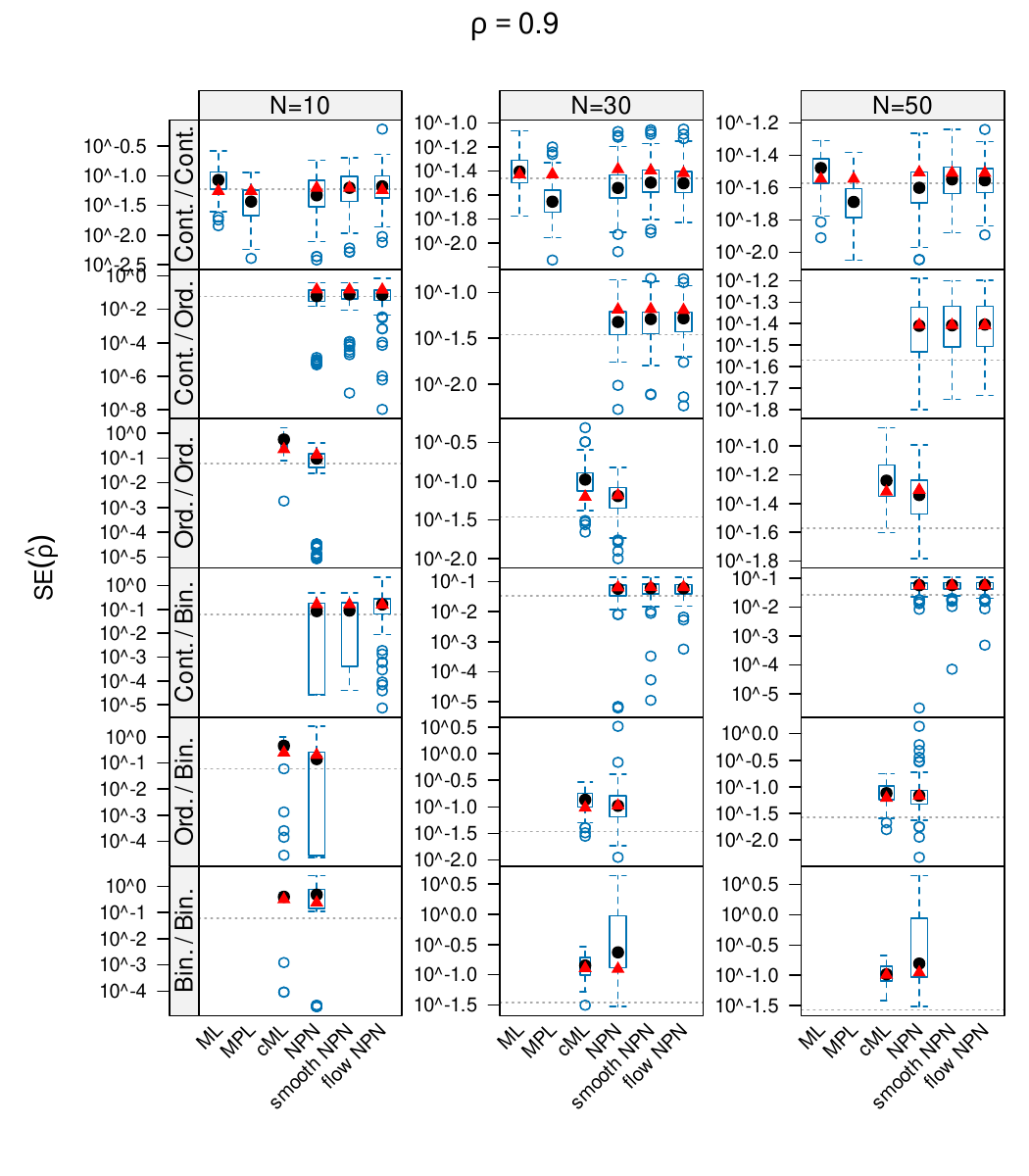} 
\end{knitrout}
\end{center}
\caption{Bivariate polychoric correlations: Distribution of $100$ standard errors for
estimators of $\rho=0.9$.}
\end{figure}

\clearpage

\section{Multivariate Polychoric Correlations} \label{app:mpoly}

Simulation results for multivariate ($J > 2$) polychoric correlations.

\begin{figure}
\begin{knitrout}
\definecolor{shadecolor}{rgb}{0.969, 0.969, 0.969}\color{fgcolor}
\includegraphics[width=\maxwidth]{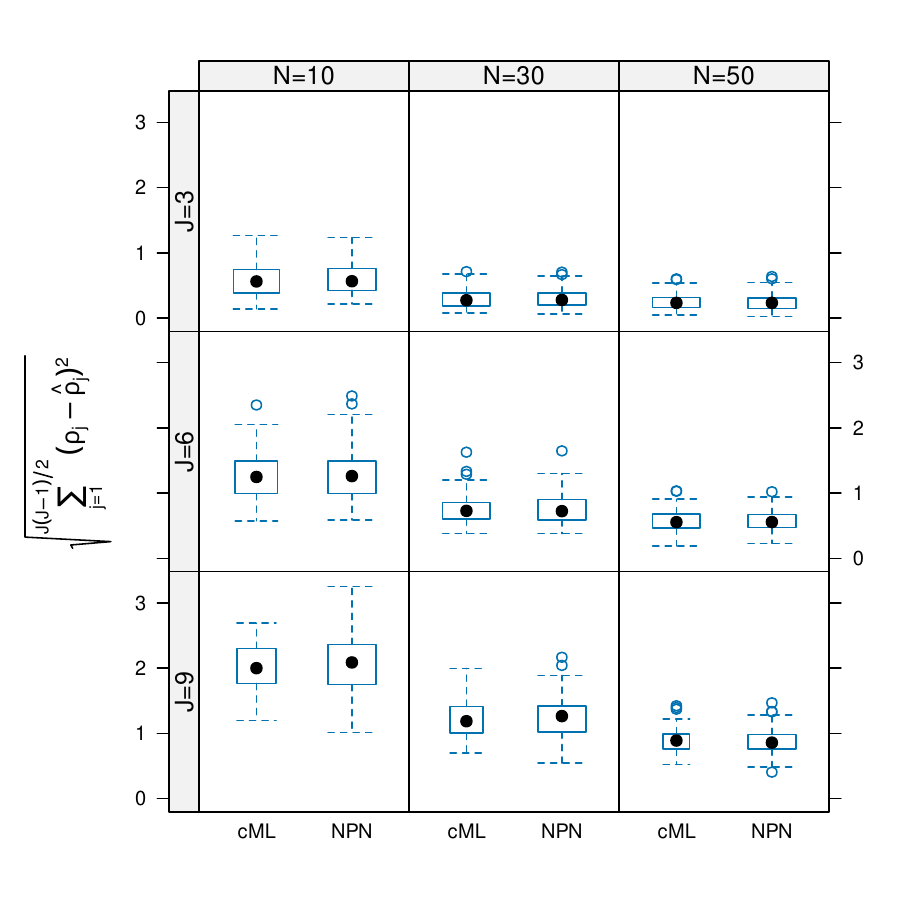} 
\end{knitrout}
\caption{Multivariate polychoric correlations: Distribution of $100$
root-mean squared standard errors for estimators of polychoric correlations
$\rho$ maximising the composite (cML, implemented in \code{mvord::mvord}) and nonparanormal (NPN, implemented in \code{tram::Mmlt})
log-likelihoods. The width of the boxes is proportional to the square root
of the number of successful optimisations out of $100$ runs.}
\end{figure}

\begin{figure}
\begin{knitrout}
\definecolor{shadecolor}{rgb}{0.969, 0.969, 0.969}\color{fgcolor}
\includegraphics[width=\maxwidth]{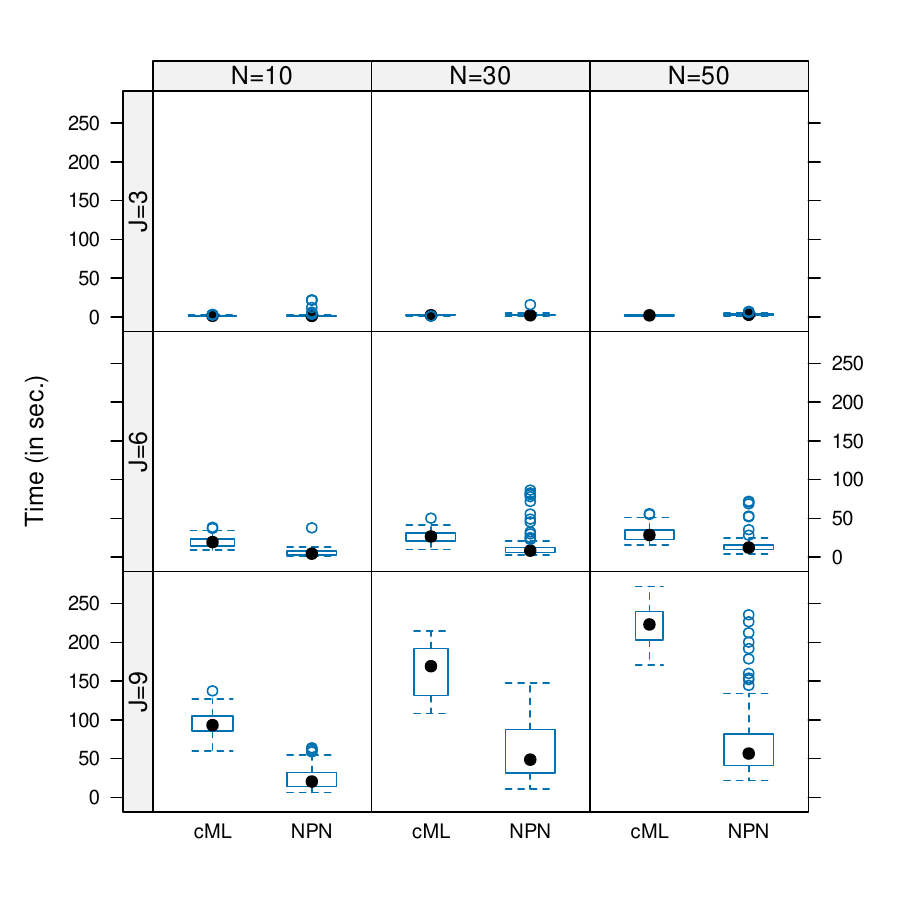} 
\end{knitrout}
\caption{Multivariate polychoric correlations: Distribution of running $100$ times 
when estimating polychoric correlations
maximising the composite (cML, implemented in \code{mvord::mvord}) and nonparanormal
(NPN, implemented in \code{tram::Mmlt}) log-likelihoods. The width of the boxes is proportional to the square root
of the number of successful optimisations out of $100$ runs.}
\end{figure}

\begin{table}[h!]
\begin{tabular}{lllD{.}{.}{0}D{.}{.}{0}D{.}{.}{0}}
\toprule
& && \multicolumn{3}{c}{N}\\
\cmidrule{4-4}\cmidrule{5-5}\cmidrule{6-6}
Method&J && \multicolumn{1}{c}{N=10}&\multicolumn{1}{c}{N=30}&\multicolumn{1}{c}{N=50}\\
\midrule
cML&J=9 && 67  & 47  & 31 \\
   &J=6 && 79  & 99  & 99 \\
   &J=3 && 94  & 99  & 99 \\
NPN&J=9 && 100 & 100 & 100\\
   &J=6 && 100 & 100 & 100\\
   &J=3 && 100 & 100 & 100\\
\bottomrule
\end{tabular}

\caption{Multivariate polychoric correlations: Number of successful optimisations in $100$ simulation runs
optimising composite (cML, implemented in \code{mvord::mvord}) and nonparanormal
(NPN, implemented in \code{tram::Mmlt}) log-likelihoods.}
\end{table}
 \end{appendix}

\clearpage

\end{document}